\documentclass[aps,prb,twocolumn,amsmath,amssymb,superscriptaddress,floatfix]{revtex4-2}
\usepackage{graphicx} 
\usepackage{physics} 
\usepackage{bm}
\usepackage[usenames]{color}
\bibstyle{apsrev.bib}
\usepackage{amsmath}
\usepackage{animate}
\usepackage{comment}
\usepackage{dcolumn}
\usepackage{bm}
\usepackage{tikz}
\usetikzlibrary{arrows.meta}
\usepackage{subcaption}
\usepackage{hyperref}
 \usepackage{array,multirow}
\usepackage{dcolumn}
\usepackage{tabularx}
\usepackage{ragged2e}
\usepackage{xcolor,colortbl}

\definecolor{mygrey}{rgb}{0.9, 0.9, 0.9}

\newcommand{\be}{\begin{equation}}
\newcommand{\ee}{\end{equation}}
\newcommand{\beqn}{\begin{eqnarray}}
\newcommand{\eeqn}{\end{eqnarray}}

\begin{document}

\title{Activity propagation with Hebbian learning}

\author{Will T.~Engedal}
\thanks{These authors contributed equally}
\affiliation{Department of Physics and Astronomy, Northwestern University, Evanston, Illinois 60208, USA}

\author{R\'obert Juh\'asz}
\thanks{These authors contributed equally}
\affiliation{HUN-REN Wigner Research Centre for Physics, H-1525 Budapest, P.O.Box 49, Hungary}

\author{Istv\'an A.~Kov\'acs}
\email{istvan.kovacs@northwestern.edu}
\affiliation{Department of Physics and Astronomy, Northwestern University, Evanston, Illinois 60208, USA}
\affiliation{Northwestern Institute on Complex Systems, Northwestern University, Evanston, Illinois 60208, USA}
\affiliation{Department of Engineering Sciences and Applied Mathematics, Northwestern University, Evanston, Illinois 60208, USA}
\affiliation{NSF-Simons National Institute for Theory and Mathematics in Biology, Chicago, Illinois 60611, USA}

\date{\today}

\begin{abstract} 
We investigate the impact of Hebbian learning on the contact process, a paradigmatic model for infection spreading, which has been also proposed as a simple model to capture the dynamics of inter-regional brain activity propagation as well as population spreading. Each of these contexts calls for an extension of the contact process with local learning. We introduce Hebbian learning as a positive or negative reinforcement of the activation rate between a pair of sites after each successful activation event. Learning can happen either in both directions motivated by social distancing (mutual learning model), or in only one of the directions motivated by brain and population dynamics 
(source or target learning models). 
Hebbian learning leads to a rich class of emergent behavior, where local incentives can lead to the opposite global effects. In general, positive reinforcement (increasing activation rates) leads to a loss of the active phase, while negative reinforcement (reducing activation rates) can turn the inactive phase into a globally active phase. 
In two dimensions and above, the effect of negative reinforcement
is twofold: it promotes the spreading of activity, but at the same time gives rise to the appearance of effectively immune regions, entailing the emergence of two distinct critical points. Positive reinforcement can lead to Griffiths effects with non-universal power-law scaling, through the formation of random loops of activity, a manifestation of the ``ant mill" phenomenon.
\end{abstract}

\maketitle

\section*{Introduction}

Biological and social systems show learning and adaptation across a broad range of scales \cite{ROSEN197539,Guirfa2003,turev}. Yet, understanding the impact of local learning mechanisms remains mostly unexplored even in simple models. Here, we focus on a paradigmatic model of activity (or infection) spreading, the contact process (CP) \cite{harris1974} and explore the impact of local learning on its dynamics, motivated by Hebbian learning~\cite{Hebb2005}. 
As introduced by T.~E.~Harris, the CP is a variant of the susceptible-infected-susceptible (SIS) model that is equivalent to the standard SIS model on regular lattices~\cite{harris1974}. The CP is a continuous-time Markov process that acts on a network of sites with each site either in an ``active" or ``inactive" state. Other common terminology for these states are infected/healthy and occupied/unoccupied. Each site can be active multiple times, like how a person can get sick many times in the case of the common cold or COVID, or how a neuron or brain region can be active on multiple occasions. 

In the case of infection spreading, it is natural to explore what happens if individuals learn from the dynamics and lower the rate at which they interact with those individuals that infected them, a form of negative reinforcement, or anti-Hebbian learning \cite{Magnasco2009-we}. For example, a reinforcement effect similar to social distancing \cite{teVrugt2020} would result in a mutual learning mechanism, where the two sites involved would reduce the time spent together.

In the context of brain dynamics, each site can be viewed as a small region of the brain (on the order of a mm$^3$) that can be either active or inactive and each edge can be viewed as a connection between the brain regions~\cite{MorettiMunoz2013}. 
As chemical synapses in the brain are directed, this motivates implementing a directed learning mechanism, either impacting the connection starting at the source or at the target.
Positive reinforcement between brain regions aligns with the observation that brain regions that fire together will wire together more strongly \cite{MOLDAKARIMOV2008667, PhysRevX.14.021001, Lynn2024}. 

As a further motivation, the CP has been also applied to the spreading of populations \cite{Oborny}. For example, parasites spreading on a background of hosts can induce local adaptations through a triggered immune response and resistance resulting from the host-parasite ``arms race" \cite{doi:10.1086/653002,10.1002/evl3.274,https://doi.org/10.1111/1365-2435.12012,Brown2025-ue}, leading to a form of negative reinforcement. Positive reinforcement could also happen along ecological corridors that are strips of favorable conditions (e.g.~land and water) connecting fragmented habitats \cite{https://doi.org/10.1111/j.0906-7590.2003.03620.x}. Once established, these corridors can be preferably used later by the population. 

Both negative and positive reinforcement leads to heterogeneities in the system if started from a homogeneous network, a form of kinetic disorder. Here, we address the question of how local learning changes the large-scale emergent behavior of the system, compared to the standard CP.
The standard CP has two phases, an active phase and an inactive phase, governed by the activation rate $\lambda$. In the active phase, the infection survives in an infinite system, whereas, in the inactive phase, the infection dies out and never returns---the system reaches an absorbing state. This is known as an absorbing phase transition~\cite{MarroDickman1999, Hinrichsen2000}. The transition in the standard CP serves as a representative of the robust directed percolation universality class (DP) \cite{henkel2008non, RevModPhys.76.663}, which is believed to contain any absorbing phase transition with short-range interactions, a scalar order parameter, a single absorbing state, and no extra conservation laws or symmetries~\cite{Janssen1981,Grassberger1982}. Outside of the CP, transitions of the DP class occur in a broad range of systems, including models of catalytic reactions~\cite{ZiffGulariBarshad1986}, damage spreading transitions \cite{Grassberger1995} and interface growth~\cite{TangLeschorn1992}.

At the critical activation rate of the transition, labeled $\lambda_c$, the standard CP follows a power-law behavior for long times. When starting from a single active site on an infinite $d$-dimensional lattice, the number of active sites $N$, survival probability $P$, and the
the root-mean-square distance of the active sites from the initial active site $R$ scale asymptotically as~\cite{MarroDickman1999,Hinrichsen2000}
\begin{align}
    N&\sim t^\eta
    \label{eq:scalingN}\\
    P&\sim t^{-\delta}
    \label{eq:scalingP}\\
    R&\sim t^{1/z}
    \label{eq:scalingR}
\end{align}
with critical exponents $\eta$, $\delta$, and $1/z$. The standard CP also showcases a phenomenon called duality \cite{schutz,vanderzande}. When starting from a fully-active lattice with periodic boundary conditions, the density of active sites, $\rho$, is identical to the survival probability of a spreading simulation, started from a single active site, yielding the relationship~\cite{MarroDickman1999,Hinrichsen2000}:
\begin{equation}
    \rho\sim t^{-\alpha}
    \label{eq:scalingRho}
\end{equation}
with $\alpha=\delta$.
Duality also implies that the critical infection rate $\lambda_c$ is the same for the two initial conditions.
Note that the critical exponents above are not independent, as they are related by the hyperscaling relation $\eta=\frac{d}{z}-\delta-\alpha$, where $d$ is the dimension of the underlying lattice~\cite{MarroDickman1999,Hinrichsen2000}.

The rest of the paper is organized as follows. The contact process with Hebbian learning is introduced in section~\ref{sec:model}. The impact of Hebbian learning is discussed in section~\ref{sec:1D} in 1D and in section~\ref{sec:2D} in 2D and above, followed by a Discussion. Detailed analytic and numerical results in 1D are presented in the Appendix~\ref{sec:analytic} for the imbalance variable, with further results for positive reinforcement in Appendix~\ref{sec:pos}.

\section{Model and Methods}
\label{sec:model}
\begin{figure}
    \centering
    \begin{subfigure}{0.45\columnwidth}
        \centering
        \subcaption{initial state}
        \vskip-5mm
        \includegraphics[]{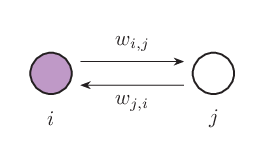}
    \end{subfigure}
    \begin{subfigure}{0.45\columnwidth}
        \centering
        \subcaption{mutual learning}
         \vskip-5mm
        \includegraphics[]{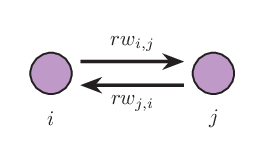}
    \end{subfigure}
    \vskip -5mm
    \begin{subfigure}{0.45\columnwidth}
        \centering
        \subcaption{source learning}
         \vskip-5mm
        \includegraphics[]{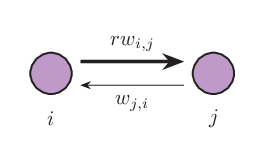}
    \end{subfigure}
    \begin{subfigure}{0.45\columnwidth}
        \centering
        \subcaption{target learning}
         \vskip-5mm
        \includegraphics[]{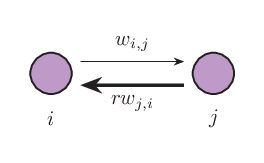}
    \end{subfigure}
    \vskip-8mm
\caption{\justifying\textbf{Illustration of our Hebbian learning rules in the contact process.} (a) denotes the initial condition while (b), (c), and (d) denote successful activation for the mutual learning, target learning, and source learning models, respectively. Purple denotes an active site and bold denotes a modified connection, multiplied by the reinforcement parameter, $r$.}
\label{fig:Choices}
\end{figure}

\begin{center}
\begin{table}[t]
\caption{\justifying \label{tab:summary} \textbf{Summary of the studied learning models.}
The precise value of the reinforcement parameter $r$ is often irrelevant, apart from whether it corresponds to negative ($r<1$) or positive ($r>1$) reinforcement. Gray background color indicates the lack of a phase transition. For $r<1$, the behavior can depend on the 
initial condition being a fully active lattice (full) or a single active site (spreading). We use the following abbreviations: DyP: dynamical percolation, GP: Griffiths phase. LDP stands for ``learning" DP, where certain aspects are governed by the DP universality class, see text. (*) 
stands for emergent quenched directional configuration(s) in the $r\to0$ limit, see text.
}
\vskip -3mm
\begin{tabularx}{\linewidth}
{>{\hsize=0.20\hsize \centering \arraybackslash}X
>{\hsize=1.00\hsize \centering \arraybackslash}X
>{\hsize=1.00\hsize \centering \arraybackslash}X
>{\hsize=1.00\hsize \centering \arraybackslash}X
>{\hsize=1.00\hsize \centering \arraybackslash}X
}
\cline{1-5}
\noalign{\vskip\doublerulesep   
         \vskip-\arrayrulewidth}
\cline{1-5}
\noalign{\vskip 2pt}         
& \multicolumn{2}{c}{1D} & \multicolumn{2}{c}{2D} \\ 
      & $r<1$ & $r>1$ & $r\to0$ & $r\to\infty$ \\
\noalign{\vskip\doublerulesep}
\cline{1-5}
\noalign{\vskip 2pt}
\parbox[t]{2mm}{\multirow{3}{*}{\rotatebox[origin=c]{90}{source}}}  &    & \cellcolor{mygrey}   & full: LDP   & \cellcolor{mygrey} GP\\
& LDP  & \cellcolor{mygrey}  directional lock  & spreading: DyP & \cellcolor{mygrey} (exp. localization)\\ 
\noalign{\vskip 2pt}
\cline{1-5}
\noalign{\vskip 2pt}
\parbox[t]{2mm}{\multirow{3}{*}{\rotatebox[origin=c]{90}{mutual}}}  & full: LDP & \cellcolor{mygrey}     & full: LDP & \cellcolor{mygrey} \\ 
& spreading: ballistic & \cellcolor{mygrey}  localization   & spreading: DyP & \cellcolor{mygrey} localization  \\ 
\noalign{\vskip 2pt}
\cline{1-5}
\noalign{\vskip 2pt}     
\parbox[t]{2mm}{\multirow{3}{*}{\rotatebox[origin=c]{90}{target}}} & \cellcolor{mygrey} ballistic front  & \cellcolor{mygrey}  \phantom{XXXXX} localization &  full: LDP  spreading: & \cellcolor{mygrey}   \phantom {localization} localization \\ 
& (*)  & \cellcolor{mygrey}     & DyP (*) &\cellcolor{mygrey}  \\ 
\noalign{\vskip 2pt}
\cline{1-5}
\noalign{\vskip\doublerulesep   
         \vskip-\arrayrulewidth}
\cline{1-5}
\end{tabularx}
\end{table}
\end{center}

The CP is controlled by an activation rate $\lambda$ and a rate $\mu$ of spontaneous deactivation, which is usually chosen to be one for convenience. In the numerical implementation, we follow the standard simulation prescription by Dickman~\cite{Dickman1999}. An active site $i$ is chosen at random from a list of $N$ active sites. The site transitions to the inactive state with probability $1/(1+\lambda)$ and attempts to activate a neighboring site with probability $\lambda/(1+\lambda)$. 
In order to approximate a continuous-time Markov chain, each of these steps increases time by $\Delta t=1/N$, so that, on average, each active site acts once per unit of time. In the standard CP, if activation is chosen, a neighboring site is chosen at random with uniform probabilities. If this site is inactive, the activation attempt is deemed successful and the neighbor transitions to the active state. 
For actual biological, social and physical networks, relationships between sites need not be uniform. To model this, we introduce edge weights, carried in a weight matrix, $W$, where $w_{i,j}$ is the weight of the connection from site $i$ to site $j$. Then, the direction of activation is chosen with probability proportional to the edge weight in that direction. As a simple example, if site $i$ in a 1D chain has chosen to activate a neighbor, it will attempt to activate site $i+1$ with probability $w_{i,i+1}/(w_{i,i-1}+w_{i,i+1})$ and it will attempt to activate site $i-1$ with probability $w_{i,i-1}/(w_{i,i-1}+w_{i,i+1})$.

To model Hebbian learning in the CP, we choose an event-based, multiplicative type of reinforcement rule: the weights are multiplied with a constant factor $r$ after each successful activation event between sites $i$ and $j$. This can be realized according to one of the three possible scenarios: mutual, source or target learning, as illustrated in Fig.~\ref{fig:Choices} and summarized in Table~\ref{tab:summary}. 
In the mutual learning model, both $w_{i,j}$ and $w_{j,i}$ are multiplied by the reinforcement parameter, $r$. In the source learning model, only the weight $w_{i,j}$ is multiplied by $r$, while in the target learning model only $w_{j,i}$ changes. 
In all cases, if site $i$ attempts to activate the already-active site $j$, $w_{i,j}$ and $w_{j,i}$ are left unchanged.
In line with the usual intuitive picture of Hebbian learning, reinforcement happens in these models only at events when two sites become simultaneously active. However, simultaneous activity alone does not lead to reinforcement, as a direct activation event is needed.

Note that even with the application of a reinforcement rule, the total rate of activations sourcing from a given site is invariant ($\lambda$), as changes in the weights only affect the preferences according to which a site spends it's contact time. In other words, we have an implicit normalization for the total activation rate and learning induces directional selectivity towards or against some partners. Without some sort of normalization, Hebbian learning is in general unstable as either all activation rates grow until a maximum allowed value or all rates decay to zero \cite{Miller}.

Our implementation of local learning is in stark contrast to past attempts in the literature to capture adaptation in spreading processes. In the traditional framework of adaptive networks \cite{PhysRevLett.96.208701,adaptive}, the topology of the connections is changing, while in the more recent approach of ``awareness" \cite{Odor2025,PhysRevResearch.7.L012061}, the infection rate changes according to the local or global activity. Hence, our implementation of Hebbian learning is a qualitatively different adaptation mechanism that might in practice appear simultaneously with other modes of adaptation, such as rewiring the topology or adjusting the local infection rates based on awareness.

To investigate the effects of our Hebbian learning rules on the CP, we consider two initial conditions---spreading from a single active site or decaying from a fully-active lattice.  For spreading simulations, we monitor the number of active sites $N$, the survival probability $P$, and the mean square radius $R$; while for a fully active initial condition we monitor the density of active sites $\rho$. 
Initially, $w_{i,j}=1$ for all neighbors $(i,j)$. 
Each edge weight can then be written as $w_{i,j}=r^{u_{i,j}}$, where $u_{i,j}$ is the number of times the connection $w_{i,j}$ has been reinforced. 
In the 1D chain case, the quotient $w_{i,i-1}/(w_{i,i-1}+w_{i,i+1})$ can be written as $1/(1+r^{n_i})$ where $n_i=u_{i,i+1}-u_{i,i-1}$ is the difference in usage between outgoing edges. Likewise, $w_{i,i+1}/(w_{i,i-1}+w_{i,i+1})$ can be written as $1/(1+r^{-n_i})$. Hence, in the 1D chain, all relevant information concerning the actions of site $i$ can be stored in the quantity $n_i$, which we refer to as the \emph{imbalance} of the site. With this definition, the imbalance of a site is zero if it has activated both of its neighbors the same number of times, indicating no bias in the preferences for future infection events.

We note that the contact process endowed with the above event-based learning rules is still a Markov process, however, on an extended state space: besides the usual occupation variables of the standard CP, the configurations of the system are also characterized by the additional set $\{u_{i,j}\}$ (or $\{n_i\}$ in 1D) which encode the actual weights. This model differs from the standard CP in that it has an infinite number of absorbing configurations differing only in the weight variables. As a consequence, our model is exempted from the scope of the DP conjecture mentioned in the Introduction. 

Note that to check the emergent behavior for values of $r<1$, it is also useful to investigate the limiting case $r\to 0$. When tracking the number of times each edge has been used, as $r$ approaches 0, the site will choose to spread activity along the least used edge more often. In the $r\to 0$ limit, 
activity always spreads along the least-used edges. If there are more than one of such edges, the direction of activity spread is chosen uniformly randomly, like in the standard CP. Similarly, for $r>1$ it is useful to consider the deterministic case with $r\to\infty$, where activity spreads along the most-used edges. 

For investigating spreading from a single active site, we chose our lattice size such that the spread never touches the boundary. 
For example, to run simulations to $t=10^6$ for mutual learning in 1D, it was sufficient to have lattice sizes of 
$L=10^6$ 
for $r\to 0$. For the same model in 2D, $L=2.5\,10^4$ was sufficient up to time $t=10^5$.
We typically collected over $10^4$ samples for spreading simulations and around $10^2$ for fully infected simulations.  
In 1D, fully infected simulations had a typical linear size of $L=10^6$, and in 2D $L=5\,10^3$, but at least $L=2\,10^3$ to reduce finite-size effects.

\section{Negative Reinforcement in one dimension}
\label{sec:1D}

\begin{figure*}[ht!]
    \subfloat{%
        \includegraphics[width=.33\linewidth]{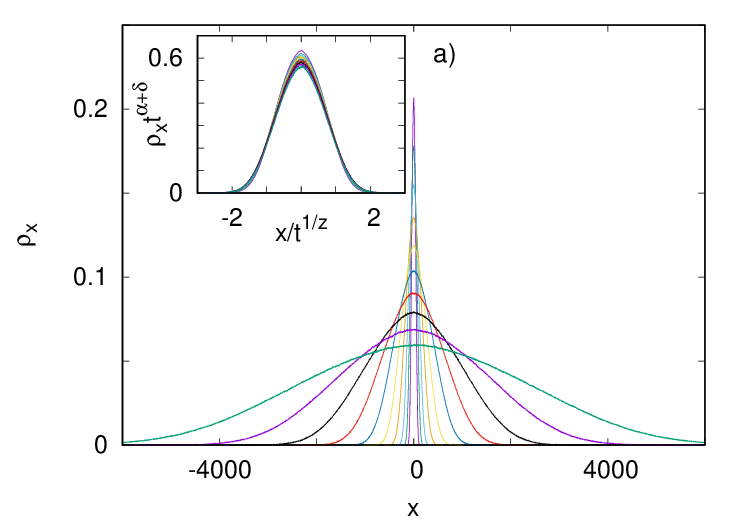}%
        \label{subfig_prof1}%
    }\hfill
    \subfloat{%
        \includegraphics[width=.33\linewidth]{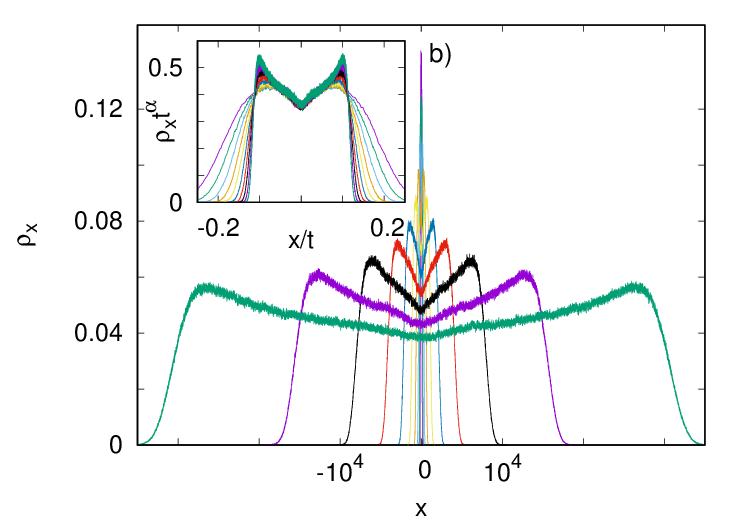}%
        \label{subfig:prof2}%
    }\hfill
    \subfloat{%
        \includegraphics[width=.33\linewidth]{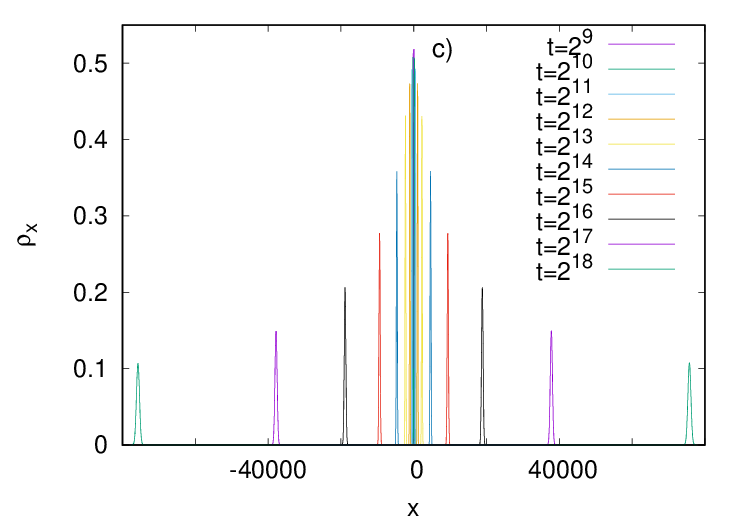}%
        \label{subfig:prof3}%
    }
\vskip-3mm
    \caption{\justifying\textbf{Density profiles of the three different learning models in one dimension.} Here, we illustrate the $\rho_x$ activity density profiles at various times starting from a single active site in the center ($x=0$) for the (a) source  ($r=0.8$), b) mutual ($r=0.5$), and c) target ($r=0.8$) learning models. 
    The source and mutual learning models are at their critical points, see Table \ref{tab:exps}, while for the target learning $\lambda=5$. The insets show a good data collapse of the profiles with the critical exponents  listed in Table~\ref{tab:exps}.
    }
    \label{fig:profiles}
\end{figure*}

\begin{table}[ht!]
\begin{center}
\caption{\justifying\textbf{Critical scaling in one dimension.} Critical activation rate and critical exponents at various negative reinforcement rates ($r<1$). Results on standard directed percolation (DP) 
are shown up to four digits after the decimal.}
\vskip -3mm
\begin{tabular}{ ccccccc }
\hline
\hline
 &$r$ & $\lambda_c$ & $\alpha$ &  $\delta$  & $\eta$  &     $z$ \\ 
DP \cite{JensenDickman1993, Jensen1999} &1.0 & 3.2979 & 0.1595 & 0.1595 & 0.3137 &  1.5808 \\ \hline
\noalign{\vskip 2pt}
 \parbox[t]{2mm}{\multirow{4}{*}{\rotatebox[origin=c]{90}{source}}} &0.9  &  3.512(1) &  0.15(1)  & 0.05(1) &  0.45(1)   &  1.54(3) \\
&0.8  &  3.721(1) &  0.14(1)  &  0.04(1) &  0.46(2)   &  1.54(2) \\
&0.5  &  4.425(2)  & 0.15(1)  &  0.03(2)  & 0.48(2)   &  1.53(4) \\
&0  &  8.13(1) & 0.16(1) & 0.04(2) & 0.45(2)  &  1.55(3) \\ \hline
\noalign{\vskip 2pt}
 \parbox[t]{2mm}{\multirow{4}{*}{\rotatebox[origin=c]{90}{mutual}}} &0.9 & 3.269(1)  & 0.29(7) & 0.00(1) & 0.73(4) & 1.005(3) \\
&0.8 & 3.248(1)  & 0.30(8) & 0.00(1) & 0.75(3) & 1.002(4) \\
&0.5 & 3.203(1) & 0.25(7) & 0.00(1) & 0.79(4) & 0.998(7) \\
&0 & 3.193(1)  & 0.17(2) & 0.00(1) & 0.83(3) & 1.00(2) \\
\hline
\hline
\end{tabular}
\label{tab:exps}
\end{center}
\end{table}

In one dimension, negative reinforcement ($r<1$) results in qualitatively different behavior in the three different learning modes as illustrated in Fig.~\ref{fig:profiles}. Here, we show the density profiles at various times starting from a single active site at the center. In the source learning model, the activity peaks at the center and spreads according to the DP dynamical scaling, 
as illustrated at the critical point in Table~\ref{tab:exps}. 
We find that even the scaling function of density profiles---plotted in the inset of Fig.~\ref{fig:profiles}---has the same non-Gaussian tail, $\tilde\rho(x)\sim e^{-Cx^{\Phi}}$ with $\Phi\approx2.4$ as observed for the standard CP \cite{henkel2008non}. 
The critical mutual learning model has a qualitatively different behavior: the activity peaks at two ballistically advancing front regions.
In stark contrast, there is no active bulk phase in the target learning model, with only two narrow fronts propagating outwards ballistically that have a finite lifetime.

In the following, we present detailed scaling results for the critical source and mutual learning models, supported by analytic arguments and further measurements in the Appendix~\ref{sec:analytic} on the behavior of the imbalance variable. For the target learning model, there is no stationary state with a positive density of active sites, no matter how large $\lambda$ is, as discussed in the Appendix~\ref{sec:pos}.

\subsection{Source learning model}

\begin{figure*}[ht!]
    \subfloat{%
        \includegraphics[width=.4\linewidth]{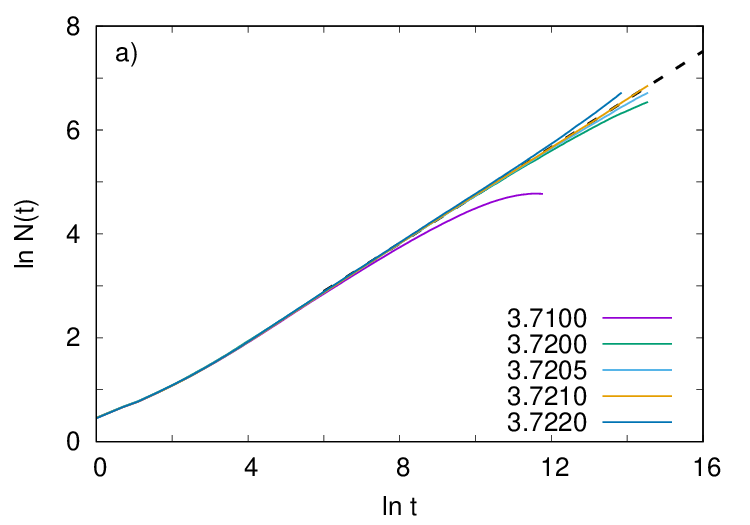}%
        \label{subfig:DAlpha0.9Nvt}%
    }
    \subfloat{%
        \includegraphics[width=.4\linewidth]{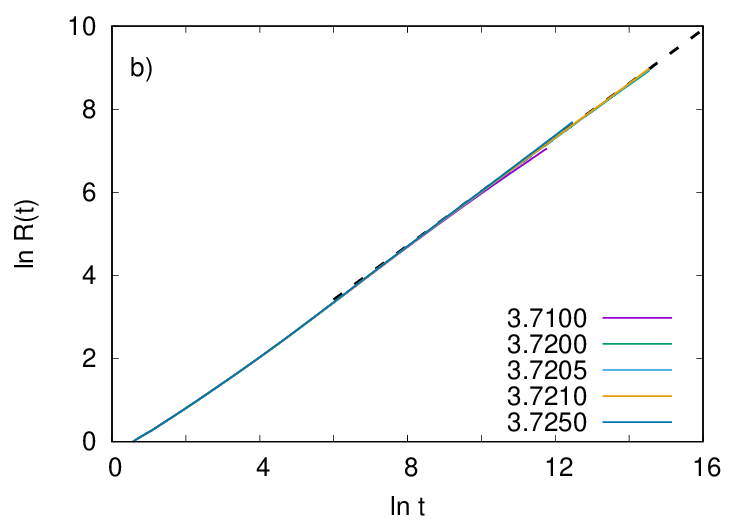}%
        \label{subfig:DAlpha0.9Rvt}%
    }\\
    \subfloat{%
        \includegraphics[width=.4\linewidth]{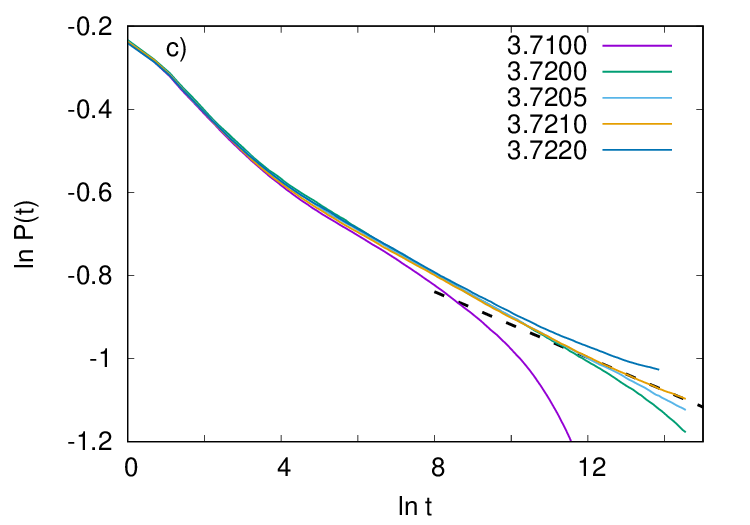}%
        \label{subfig:DAlpha0.9Pvt}%
    }
    \subfloat{%
        \includegraphics[width=.4\linewidth]{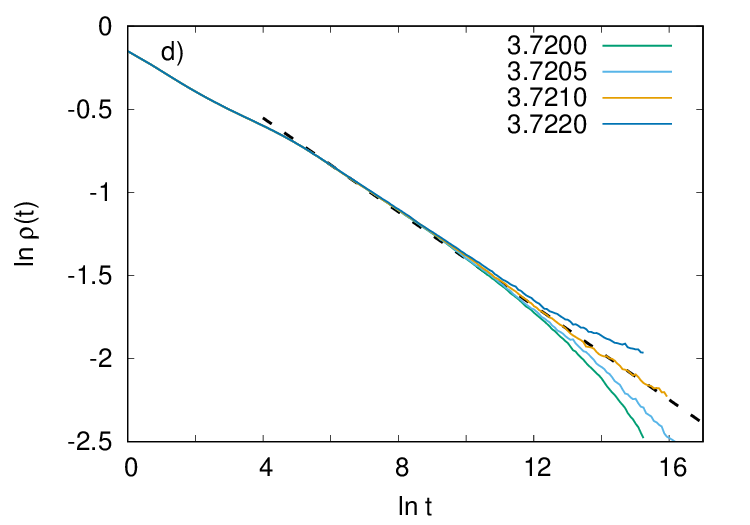}%
        \label{subfig:DAlpha0.9Rhovt}%
    }
    \vskip-3mm
    \caption{\justifying\textbf{Critical scaling in the source learning model with negative reinforcement.} Here, we illustrate a) the number of active sites $N$, b) the mean square radius $R$, c) the survival probability $P$, and d) the density of active sites $\rho$ at $r=0.8$. The asymptotic critical behavior is illustrated by the straight dashed line. 
    }
    \label{fig:DAlpha0.8}
\end{figure*}

We have explored several values of the learning rate $r$, leading to qualitatively similar results for all $r<1$ values. Here, as an illustration, we focus on presenting the results at $r=0.8$, shown in Fig.~\ref{fig:DAlpha0.8}, with additional results summarized in Table~\ref{tab:exps}.
We find that the critical point of the standard CP is inactive with negative reinforcement, and the critical point is monotonically shifted towards larger values of $\lambda$ as $r$ decreases.
In terms of standard observables (Fig.~\ref{fig:DAlpha0.8}), all $r<1$ cases are similar in the sense that within the error they cannot be distinguished from the DP universality in terms of $\alpha$, $z$, and $\eta+\delta$. Following directly from the definitions, $\eta+\delta$ governs the total number of active sites in the surviving runs as $N_s\sim t^{\eta+\delta}$. 
This means that the observables conditioned on survival, $N_s$ and $R(t)$ show DP critical behavior, but the survival probability itself is affected by reinforcement, see the altered values of $\delta$ in Table~\ref{tab:exps}. This entails the change of $\eta$, characterizing the number of active sites in all trials. 
The latter two exponents seem to change hardly with the reinforcement parameter. 
Such a scenario is not unprecedented for the CP: the surface critical behavior is also known to be characterized by a modified order parameter exponent, while the exponents related to observables conditioned on survival are unaffected \cite{henkel2008non}. 
Note that duality is violated in this model, as evident from the different values of $\alpha$ and $\delta$, although the transition happens at the same $\lambda_c$ for both initial conditions.
We refer to 
this type of critical behavior with certain exponents governed by DP, while others (here $\delta$ and $\eta$) deviating from DP, 
as a manifestation of the learning DP (LDP) behavior.

A deeper understanding of this LDP behavior can be obtained through the imbalance variable. As introduced earlier, the imbalance of a site $n_i(t)$ is the signed sum of local successful infection events from site $i$, $-1$ meaning left infection, $+1$ meaning right infection. 
As shown in Fig.~\ref{fig_ndist}b, the imbalance distributions tend to a limit distribution at late times, which has Gaussian tails $p_n\sim e^{-Cn^2}$ with and $r$-dependent constant $C$ for $|n|\gg 1$.  This is found to hold irrespective of whether the model is in the active phase or at its critical point. The Gaussian form is in line with our analytic expectations based on a simple mean-field approximation presented in the Appendix~\ref{sec:analytic}.
We note that, the local variable $n_i$ at a fixed site has the same limit distribution as the ensemble of local variables, and this limit distribution is symmetric under $n\to -n$.   
The Gaussian tail of the limit distribution of imbalance variables translates to a log-normal tail of the distribution of instantaneous weights.

\subsection{Mutual learning model}

\begin{figure*}[ht!]
    \subfloat{%
        \includegraphics[width=.4\linewidth]{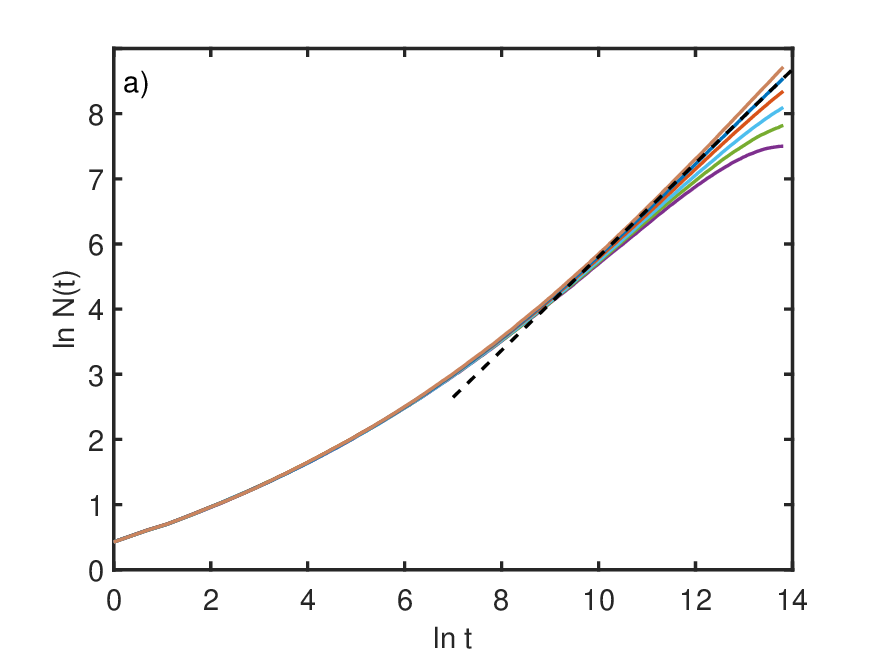}%
        \label{subfig:Alpha0.9Nvt}%
    }
    \subfloat{%
        \includegraphics[width=.4\linewidth]{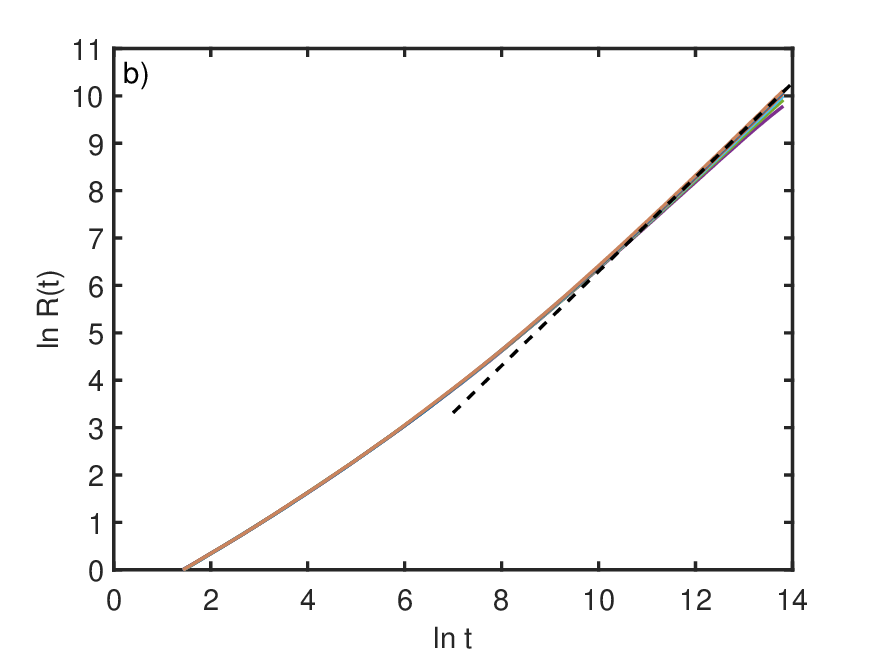}%
        \label{subfig:Alpha0.9Rvt}%
    }\\
    \subfloat{%
        \includegraphics[width=.4\linewidth]{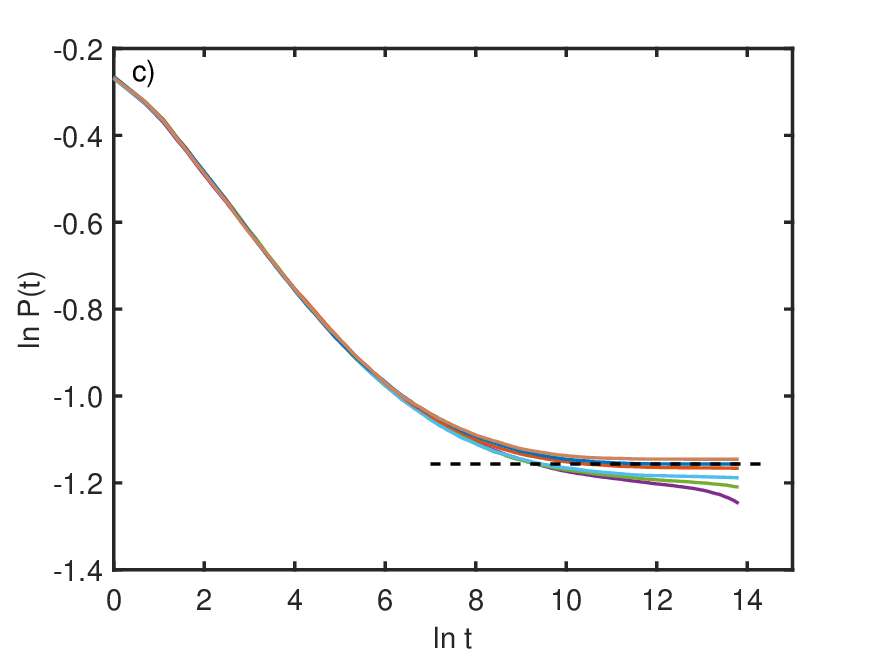}%
        \label{subfig:Alpha0.9Pvt}%
    }
    \subfloat{%
        \includegraphics[width=.4\linewidth]{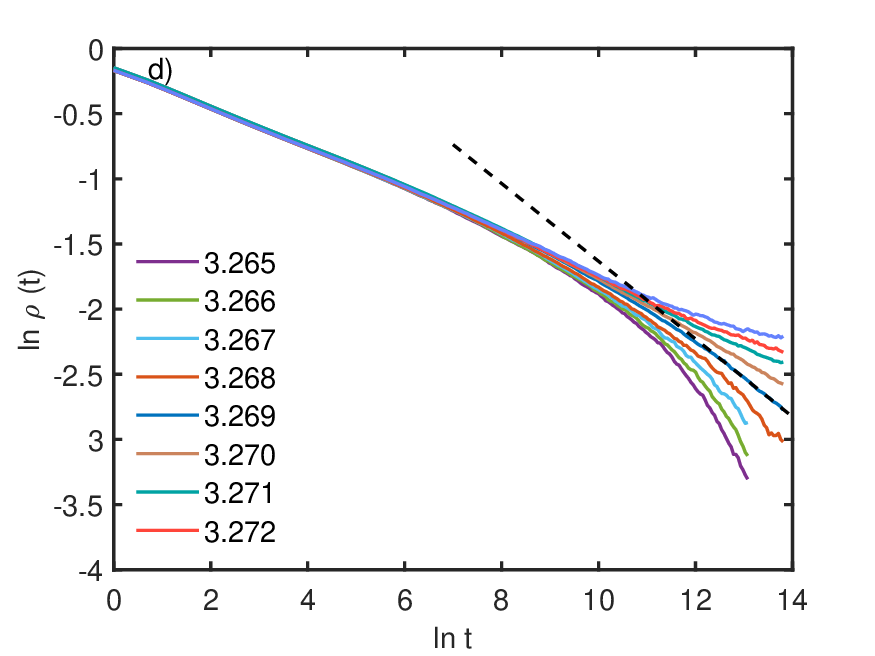}%
        \label{subfig:Alpha0.9Rhovt}%
    }
    \vskip-3mm
    \caption{\justifying\textbf{Critical scaling in the mutual learning model with negative reinforcement.} Here, we illustrate the number of a) infected sites $N$, b) the mean square radius $R$, c) the survival probability $P$, and d) the density $\rho$ for $r = 0.9$. The asymptotic behavior is illustrated by dashed lines.
    }
    \label{fig:Alpha0.9}
\end{figure*}

Just like in the source learning case, we obtained qualitatively similar results for all values of $r<1$, illustrated here through the case of $r=0.9$, see Fig.~\ref{fig:Alpha0.9}, with all results summarized in Table~\ref{tab:exps}.
As opposed to source learning, the critical activation rate weakly decreases with decreasing $r$.
Thus the $r<1$ mutual learning model is active at the critical point of the standard CP, indicating that mutual negative reinforcement promotes infection spreading.
The other remarkable difference compared to source learning is that the survival probability $P(t)$ tends to a positive limit at late times even at the critical point. This means that regarding, as usual, the limiting value of $P(t)$ as an order parameter, the transition is discontinuous, with a formally vanishing exponent, $\delta=0$. 
The other measured critical exponents also differ from those of the source learning model. 
The dynamical exponent is close to $1$ for all $r<1$, indicating a ballistic spreading of the active zone, in agreement with the density profiles in Fig.~\ref{fig:profiles}b. 
The asymptotic velocity of spreading is found to increase with decreasing $r$. 
The number of active sites $N$ shows a power-law like scaling with a much larger $\eta$ exponent than for the standard CP or the source learning model.
%
The global density $\rho(t)$, when the process starts from a fully active initial state, decays algebraically in time, and the exponent $\alpha$ is close to the DP value for not too high values of $r$, otherwise it has a somewhat increased value compared to DP. 
Duality of the DP class is clearly broken---even the corresponding critical exponents are different $\alpha\neq \delta$.
Mutual learning in one dimension thus serves as another example of an LDP, a mixture of DP-like and non-DP behavior, although it is clearly different from the behavior of the source learning model. 

The inspection of the density profiles shown in Fig.~\ref{fig:profiles}b, as well as the measured values of dynamical critical exponents lead us to the following picture of activity spreading at the critical rate. 
At a given time, the system can be divided in three parts. An intact region, which has not yet been visited by activity, a front region with a high $[O(1)]$ local density of activity, and an interior part in which sites were reactivated many times so that the imbalance distribution is close to its limiting distribution. It is easy to see that the mutual learning rule with $r<1$ facilitates activation processes from the actual front particle toward the intact region. 
When starting from a single active site, as activity spreads outwards, the edge connecting the newly active site back towards the center of the cluster is weakened. When it is this newly activated site's turn to activate others, it is significantly more likely to activate outwards, towards inactive sites. This reduces the amount of times activation is attempted but failed, promoting spreading. 

This mechanism results in a ballistic drift of the fronts, at which the process is locally supercritical. In the interior part, the process is critical and the local density decays algebraically in time. This ballistic expansion of the critical zone (c.f.~the sublinear creeping of the front $R(t)\sim t^{1/z}$ in the standard CP) ensures that the process remains surviving with a finite probability even at the critical point. Specifically, at a fixed site $x$, the local density is $O(1)$ at time $t_x\simeq x/v$, i.e.~the time at which the front sweeps through site $x$.
Here, $v$ denotes the mean asymptotic velocity of the front.
After this event, the local density decreases in time asymptotically as $\rho_x(t)\sim (t-x/v)^{-\alpha}$. Summing up these local densities in the interior part $[-vt,vt]$, we obtain an algebraic increase of the number of active sites $N(t)\sim t^{1-\alpha}$. Assuming that the bulk criticality belongs to the DP universality class, we obtain $1-\alpha=0.8405$. Numerical estimates presented in Table~\ref{tab:exps} are somewhat lower; this deviation may be attributed to corrections stemming from the front region which is found to spread out proportionally to $t^{1/z}$ with the DP dynamical exponent $z$ on top of the ballistic advancement (not shown).

As shown in Fig.~\ref{fig_ndist_un}, the distribution of the imbalance values tends to a stationary distribution with an exponential tail, $p_n\sim e^{-an}$, with $a=0.87(2)$. This scaling form is in agreement with our mean-field calculation, shown in the Appendix~\ref{sec:analytic}. Note that the exponential tail of imbalance variable corresponds to a power-law tail of the instantaneous weights.

\section{Higher dimensional systems}
\label{sec:2D}

In higher dimensions, for simplicity, we restrict our presentation to the results obtained in the limits $r\to0$ and $r\to\infty$. These limits are computationally more feasible to study, lead to less severe scaling corrections and are usually expected to be representative of the generic cases with negative ($r<1$) and  positive ($r>1$) reinforcement, respectively. As in 1D, the target learning model is an exception at $r\to0$, as in this limit the heterogeneities become frozen, leading to an emergent quenched disorder, as discussed later in this section.

\subsection{Negative reinforcement}

 \begin{table}[ht!]
\begin{center}
 \caption{\justifying \textbf{Critical scaling in 2D.} Critical activation rate and critical exponents at negative reinforcement rate $r\to0$, in comparison to the directed percolation (DP) and dynamical percolation (DyP) universality classes. }
 \vskip -3mm
 \resizebox{0.5\textwidth}{!}{%
\begin{tabular}{ ccccccc }
\hline
\hline
\noalign{\vskip 2pt}
 & $\lambda_1$ (spread) & $\lambda_2$ (full) & $\alpha$  & $\delta$ & $\eta$ & $z$\\
DP \cite{VojtaFarquharMast2009} & 1.64874(2) & 1.64874(2) &  0.4526(7) &  0.4526(7) & 0.233(6) &  1.757(8) \\
DyP \cite{Tome, PhysRevE.59.6175} & 4.66571(3) & 4.66571(3) &  0.092 &  0.092 & 0.586 &  1.1293 \\ \hline
\noalign{\vskip 2pt}
source & 1.6768(1) & 1.695(1) & 0.44(1) & 0.07(7) &   0.65(3) & 1.12(2) \\
mutual & 1.3990(2) & 1.4675(1) & 0.44(4) & 0.07(4) & 0.71(7) & 1.08(3) \\
target  & 1.5970(5) & 1.8360(5) &  0.55(4) &  0.099(4) & 0.57(4) &  1.13(2) \\
\noalign{\vskip 2pt}
\hline
\hline
\end{tabular}
}
\label{tab:2Dexps}
\end{center}
\end{table}

\begin{figure*}[ht!]
    \subfloat{%
        \includegraphics[width=0.4\linewidth]{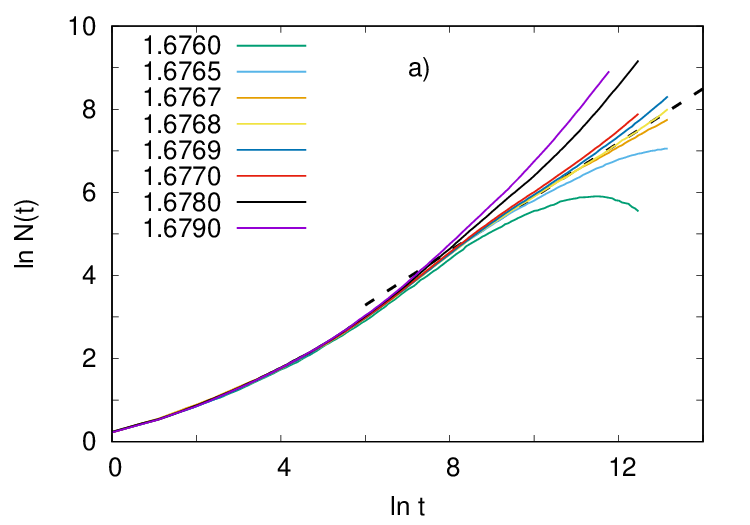}
        \label{subfig:D2DN}%
    }
     \subfloat{%
        \includegraphics[width=0.4\linewidth]{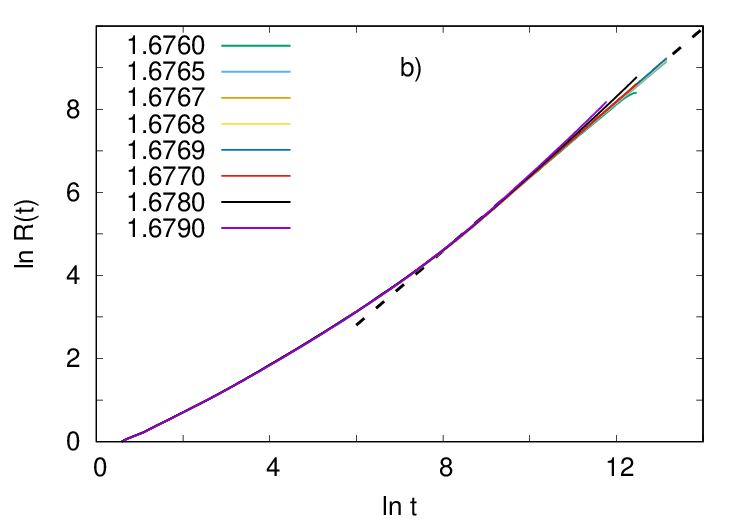}
        \label{subfig:D2DR}%
    }\\
    \subfloat{%
        \includegraphics[width=0.4\linewidth]{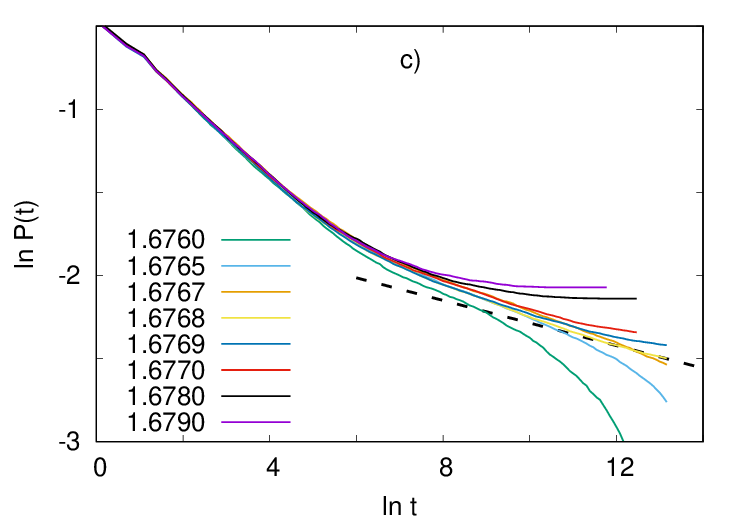}
        \label{subfig:D2DP}%
    }
    \subfloat{%
        \includegraphics[width=0.4\linewidth]{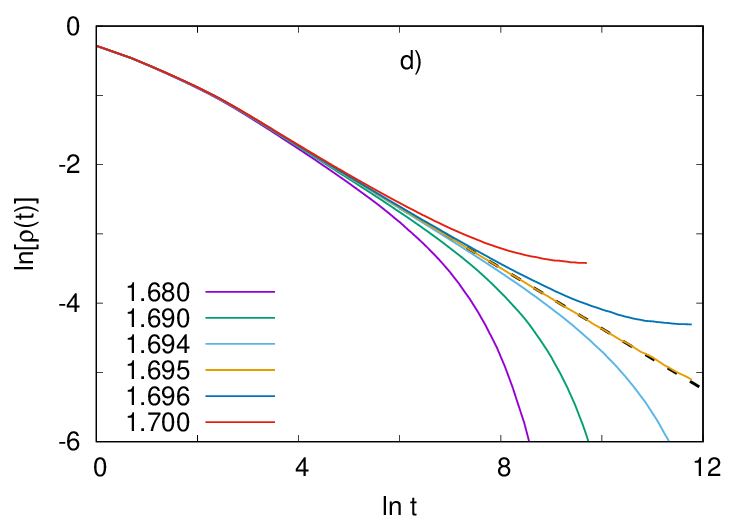}
        \label{subfig:D2DRho}%
    }
    \vskip-3mm
    \caption{\justifying\textbf{Source learning in 2D.}
    Here, we illustrate the number of a) infected sites $N$, b) the mean square radius $R$, c) the survival probability $P$, and d) the density $\rho$ for $r\to0$. 
    The asymptotic behavior is illustrated by dashed lines.}
    \label{fig:D2Dvis}
\end{figure*}

\begin{figure*}[ht!]
    \subfloat{%
        \includegraphics[width=0.4\linewidth]{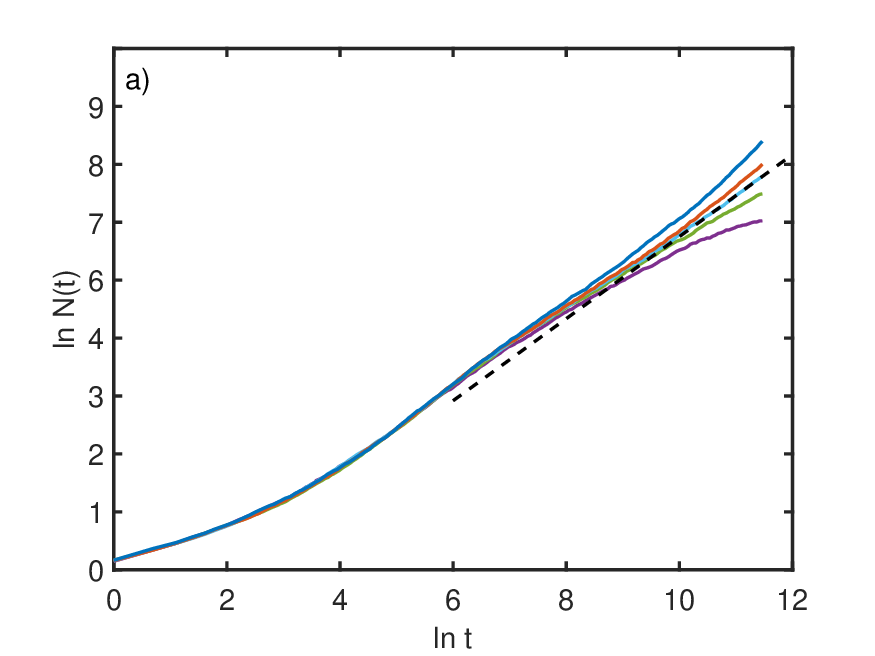}
        \label{subfig:UD2DN}%
    }
     \subfloat{%
        \includegraphics[width=0.4\linewidth]{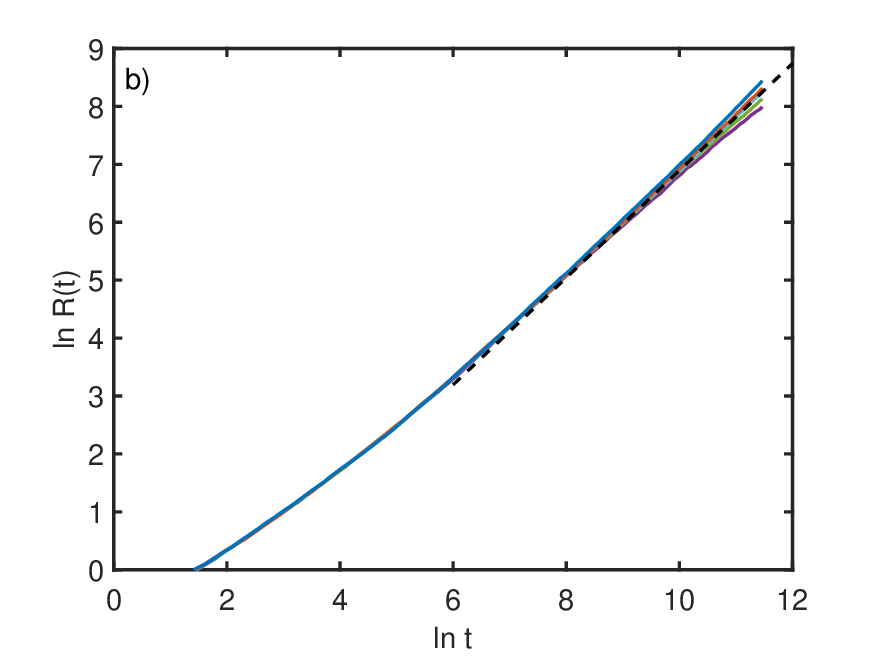}
        \label{subfig:UD2DP}%
    }\\
     \subfloat{%
        \includegraphics[width=0.4\linewidth]{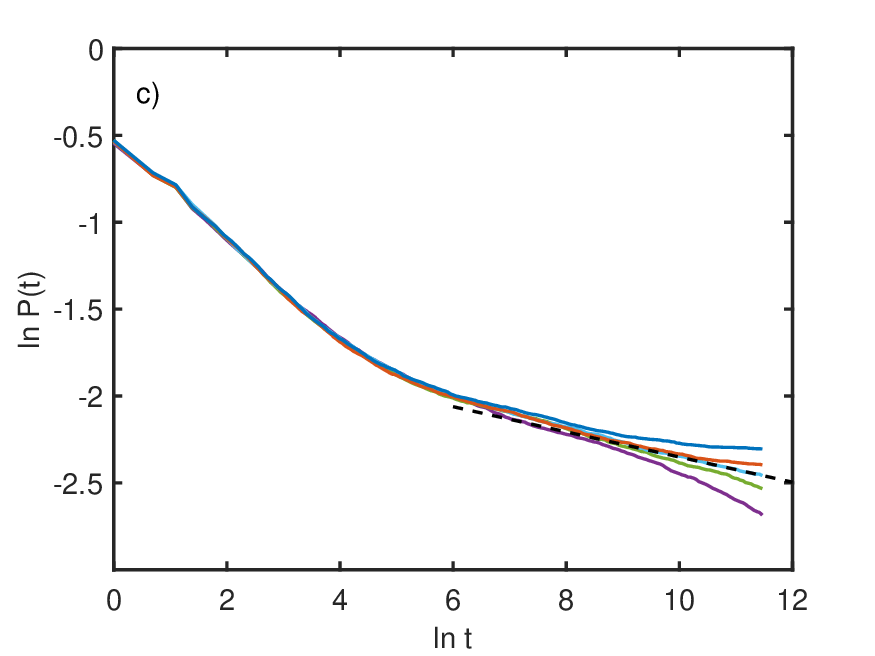}
        \label{subfig:UD2DR}%
    }
    \subfloat{%
        \includegraphics[width=0.4\linewidth]{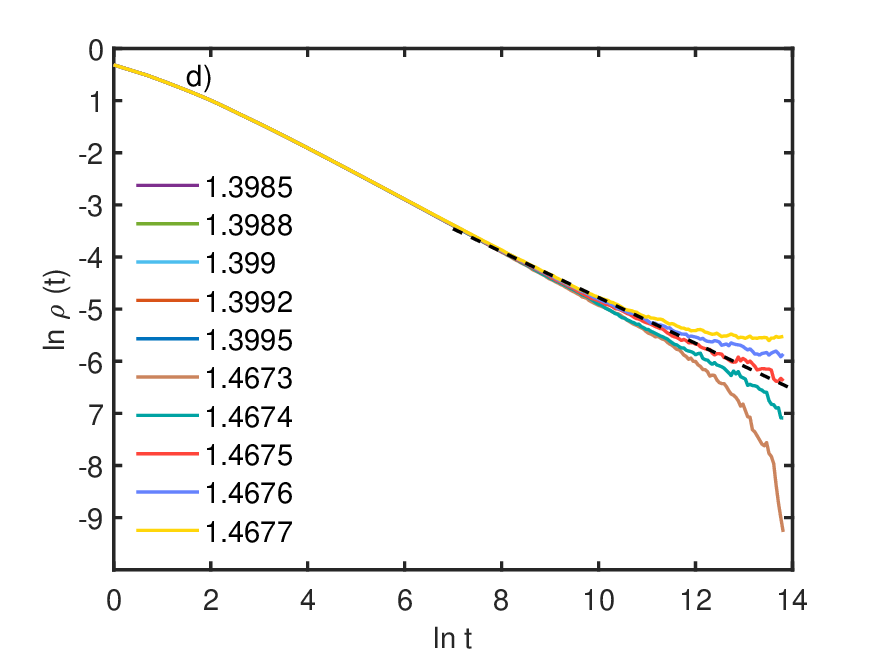}
        \label{subfig:UD2DRho}%
    }
    \vskip-3mm
    \caption{
    \textbf{Mutual learning in 2D.} The same as Fig.~\ref{fig:D2Dvis} for mutual learning.
   }
    \label{fig:UD2Dvis}
\end{figure*}

\begin{figure*}[ht!]
    \subfloat{%
       \includegraphics[width=0.4\linewidth]{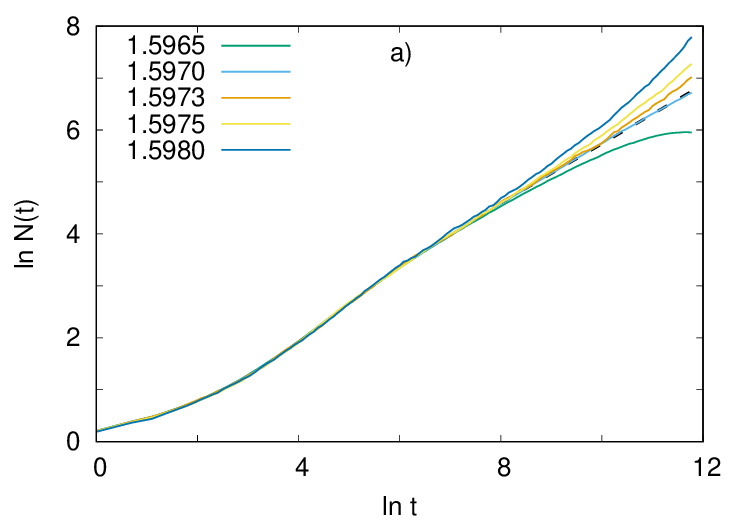}
        \label{subfig:2DTN}%
    }
     \subfloat{%
        \includegraphics[width=0.4\linewidth]{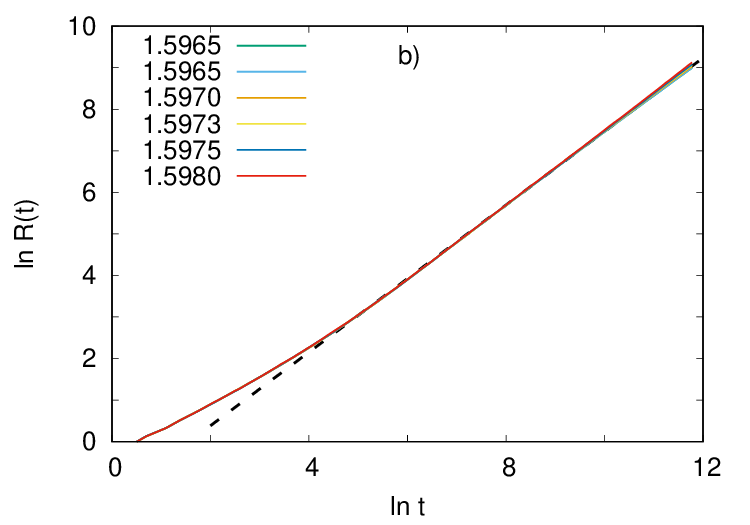}
        \label{subfig:2DTP}%
    }\\
     \subfloat{%
        \includegraphics[width=0.4\linewidth]{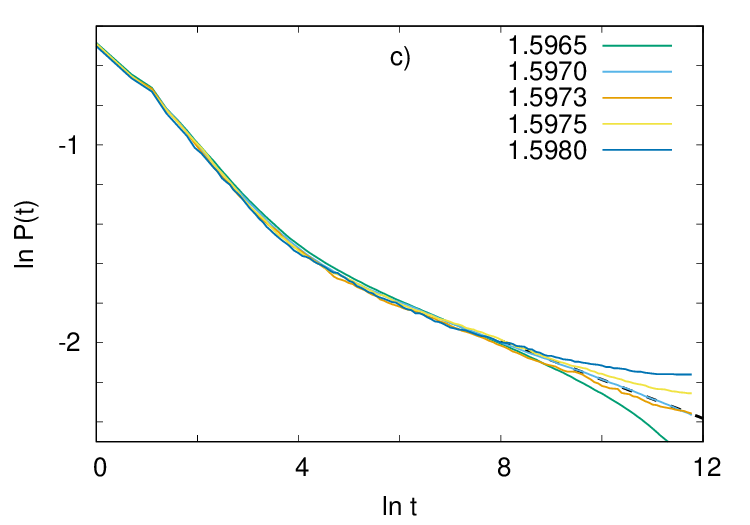}
        \label{subfig:2DTR}%
    }
    \subfloat{%
        \includegraphics[width=0.4\linewidth]{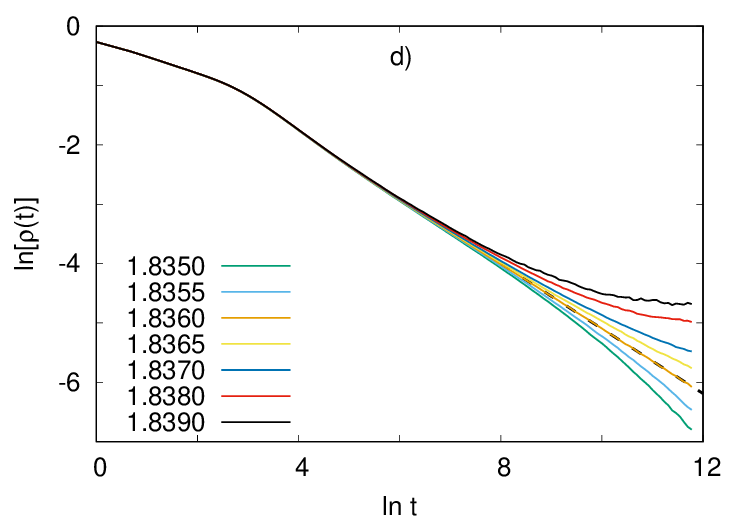}
        \label{subfig:2DTRho}%
    }
    \vskip-3mm
    \caption{
    \textbf{Target learning in 2D.} The same as Fig.~\ref{fig:D2Dvis} for target learning.
    }
    \label{fig:2DT}
\end{figure*}

\begin{figure}[ht!]
\includegraphics[width=0.8\linewidth]{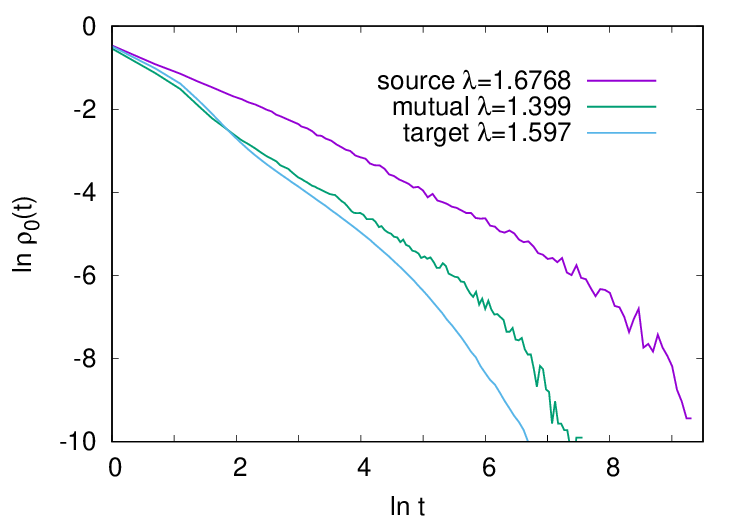}
\vskip-3mm
\caption{\justifying \textbf{Local activity scaling in 2D.} 
Time dependence of the local density $\rho_0$ at the origin ($x=0$) in the $r\to0$ limit for a critical spreading simulation (at $\lambda_1$, see Table~\ref{tab:2Dexps}) averaged for all samples for source, mutual and target learning. 
In all cases, we observe a sharp decay at long enough times, indicating a subcritical bulk phase. 
}
\label{fig_rho_local}
\end{figure}

In the source learning case, our simulation results are illustrated in Fig.~\ref{fig:D2Dvis} for $r\to0$, with the estimates of the critical exponents listed in Table~\ref{tab:2Dexps}.
For a fully-active initial condition, $\lambda_c\approx1.695$ and $\alpha$ is around the DP value, while spreading simulations indicate a lower critical rate around $\lambda_2\approx1.6768$. 
%
In the mutual learning model,  we again observe two critical points, as illustrated in Fig.~\ref{fig:UD2Dvis}, with the exponents presented in Table~\ref{tab:2Dexps}. 
Our results indicate that the system is moved to a supercritical state when run at the clean critical point but with $r<1$. Both the spreading critical point ($\lambda_1$) and the bulk critical point ($\lambda_2$) is shifted down for $r<1$ negative reinforcement, just like in the 1D case.

A similar scenario of two phase transitions is observed for the target learning model, as shown in Fig.~\ref{fig:2DT}. Here, $\lambda_1$ is below while $\lambda_2$ is above the clean critical point.
For this model with $r\to0$, a special spontaneous "freezing" of random directions takes place which is analogous to the extraordinary behavior of the same model in one dimension. 
Here, the set of least-used (smallest-weight) directions at each site constitutes the ``directional configuration" of the system.
Occupation variables then follow a simple dynamics that is completely determined by the directional configuration of the system. 
It is easy to see that there is an infinite number of directional configurations which remain invariant under the dynamics of the model. 
The only restriction for creating such invariant configurations is non-reciprocity, i.e.~whenever site $i$ has a least-used direction to site $j$ then site $j$ must not have a least-used direction to site $i$. 
A reciprocal situation is unstable against the dynamics and can only happen if the local activity ceases to exist before fixing the direction of the link.
Indicating the least-used directions by arrows on links, invariant configurations are constructed by placing arrows (in either direction) on links or not placing any, under the condition that each site has at least one outgoing arrow. 
According to our numerical observations, the $r\to0$ target learning process---when started from an isotropic directional configuration---evolves under the dynamics to an invariant directional configuration, which is chosen randomly by the fluctuating dynamics. 
Obviously, the evolution of directional configuration may also freeze locally in some domain before settling into an invariant configuration by the vanishing of activity in that domain.   
This quenching phenomenon is similar to what happens in one dimension, although the directional configuration there is trivial due to the 1D topology. 

In all cases, our observations imply the existence of two phase transitions in these models, dividing the emergent behavior into three distinct phases of i) no activity below $\lambda_1$, ii) a propagating front with no bulk activity for $\lambda_1<\lambda<\lambda_2$, and iii) an active bulk phase for $\lambda>\lambda_2$. 
%
As a further evidence for this picture, at $\lambda_1$ we eventually observe a fast decay of the local density at the origin as illustrated in Fig.~\ref{fig_rho_local}, indicating that the bulk is asymptotically inactive at $\lambda_1$. 
The $\lambda_2$ location of the bulk critical point is different in all three models, as identified by the critical point of fully infected simulations based on the scaling of the total density $\rho(t)$. 

The source and mutual learning models are close to the DP scaling at the bulk critical point, $\lambda_2$. In the case of the target learning model, for which spontaneous freezing of directions occurs (cf. the fluctuating weights in the other two models), the exponent $\alpha$ differs significantly from the DP value, although strong corrections may also not be excluded.
%
At the spreading critical point $\lambda_1$, we see a marked deviation from the DP expectations, close to the values of the dynamical percolation (DyP) universality class. In this case, target learning is closest to the DyP exponents, but the other two model variants might also show the same behavior (as suggested by $z$ and $\delta$), albeit with stronger corrections (mostly in the case of $\eta$).
The DyP is known to capture the phase transition in the susceptible-infected-recovered (SIR) model. Note that the DyP exponents satisfy a modified hyperscaling relation \cite{henkel2008non}, given by $\eta=d/z-2\delta-1$. This hyperscaling relation is also satisfied by our exponents for each of the three learning models at $\lambda_1$.
Next, we argue that indeed, at $\lambda_1$, asymptotically all three learning models are expected to follow a SIR dynamics.

As in the 1D model with mutual learning, we can distinguish three different domains of the spreading system above $\lambda_1$. These are the i) ballistically expanding annular front region, where the activity is high, ii) the unvisited domain outside of the front, and iii) the interior part which has already been swept through by the front. Spreading of the front region toward the outer empty region where there is no directional bias is facilitated compared to spreading in the inner domain, in which there is a random directional field for transmitting activity. As a consequence, for a not too high value of $\lambda$, the process is still locally subcritical in the interior, although the front region is steadily expanding outwards. Note that once activity visited a local region, the homogeneity of weights there is irreversibly broken.  
This suggests an analogy with the dynamics of the SIR model, which differs from the standard CP in that active sites do not return to the inactive state but get removed irrevocably from the system. 
The analogy is salient: the unvisited (homogeneous) local regions correspond to the susceptible state of the SIR model, whereas the visited (inhomogeneous) local regions correspond to the recovered state. 
This correspondence is obviously not perfect: in the SIR model, recovered sites can never be reactivated, while in our model this rule is less strict, since activity is able to penetrate into an inhomogeneous region. Nevertheless, as this latter region is subcritical, the penetration length is finite. Hence, concerning large-scale properties at the critical point, this circumstance is irrelevant. Based on this observation, we expect the lower (spreading, $\lambda_1$) critical points of our two-dimensional learning models to belong to the DyP universality class of the SIR model.  

Thus, the lower phase transition of the model ($\lambda_1$) is governed by the criticality of the front region, while the interior of the system is subcritical at this point.
It is clear that such a situation can be realized only in dimensions higher than $1$, in which the boundary between the intact and visited domain has a dimension $d-1>0$. On the contrary, in one dimension the criticality of the system cannot be sustained solely by a (zero-dimensional) front. As a consequence, in 1D models, the entire inner region must be critical at the spreading transition. This is in agreement with our numerical observations that our models in one dimension display only a single phase transition.


\subsection{Positive reinforcement}

\begin{figure}[ht!]
\includegraphics[width=0.8\linewidth]{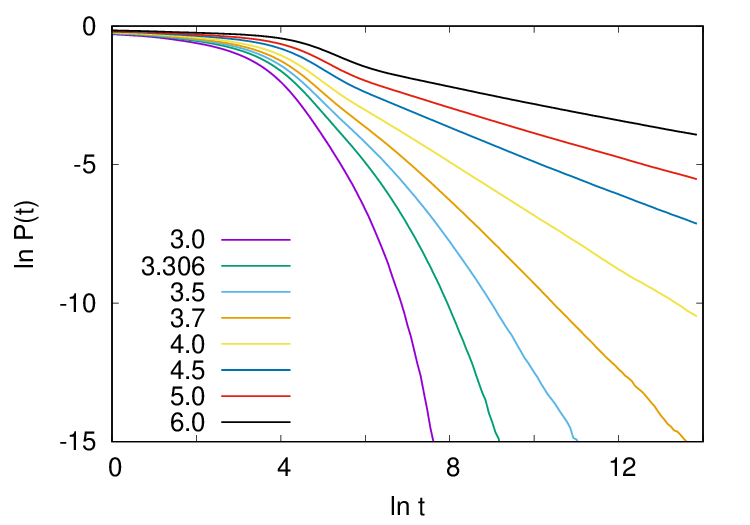}
\vskip-3mm
\caption{\justifying \textbf{Griffiths phase in the 2D source learning model with positive reinforcement.} 
In this representation, straight lines indicate a Griffiths phase of non-universal power laws, expected to appear above the critical point of the one-dimensional totally asymmetric CP, $\lambda^*=3.306(2)$ \cite{tretyakov}. Here, the Griffiths phase is a manifestation of the ``ant mill" phenomenon, where the activity propagates along randomly formed loops until it ceases to exist \cite{Delsuc}.
}
\label{fig_GP}
\end{figure}

As found in Appendix~\ref{sec:pos}, in the one-dimensional case within a mean-field treatment, there is generally no active phase with positive reinforcement.
This is also the case in higher dimensions. 
While the mutual and target learning models lead to a quick  localization of the activity for all $r>1$, the source learning case can show a qualitatively different behavior in higher dimensions. This generic phenomenon is easier to understand for large $r\to\infty$.
In this case, it is easy to see that the direction of the first successful activation from a given site "freezes" i.e. any further activation from that site can occur only in that direction. This means that learning leads to an emergent, correlated quenched disorder in the system.  
As a consequence, in spreading simulations, a train of particles is forming, the front particle performing a random walk until it revisits a site that was activated before. In any dimensions, this happens with probability one---after a finite transient---and then the activity can only propagate around the one-dimensional loop. This is a manifestation of the ``ant mill" phenomenon \cite{Delsuc}. An ant mill emerges when certain 
ants reinforce their trajectories by pheromones following each other until they die of exhaustion when accidentally forming a loop.
The simplest mathematical model considers multiplicative reinforcement in the form of a directed edge reinforced random walk, with rigorous results in all $d\geq2$ dimensions \cite{Erhard2022}.

Large loops happen with an exponentially small probability, but they are active for an exponential amount of time above the critical point of the one-dimensional totally asymmetric CP with $\lambda^*=3.306(2)$ \cite{tretyakov}. Hence, above $\lambda^*$ we observe a non-universal power-law decay of the survival probability as shown in Fig.~\ref{fig_GP} in the large $r$ limit, a hallmark of a Griffiths phase \cite{IgloiMonthus2005, Igloi2018}. The emergence of a Griffiths phase is expected to be a generic result that holds for all $r>1$ reinforcement rates in all $d\geq2$ dimensions above $\lambda^*$.
Note that in one dimension, loops have only two sites, so a Griffiths phase with the same mechanism cannot occur.

\section*{Discussion}

In this study, we introduced generalizations of the contact process with a local learning rule inspired by Hebbian learning, leading to rich emergent phenomena. We found that local incentives can lead to the opposite global effects, like locally avoiding previously activating or activated neighbors can turn the inactive phase into a globally active phase, and vice versa. 
For example, positive reinforcement in the mutual learning model is akin to quarantining. In this case, if infection is passed between two sites, they preferentially interact with each other, instead of  their other neighbors, leading to activity localization at a finite scale that depends on the value of the reinforcement parameter, $r>1$. such localization hinders activity spreading at the global scale, forcing the system to be in the inactive phase.

In stark contrast to standard critical phenomena, microscopic differences in the learning models can lead to dramatic changes in the emergent behavior, from completely destroying a phase transition to introducing an additional phase transition. For example, in two dimensions, with negative reinforcement, we see different phase transition depending on the initial conditions.
At the fully infected (bulk) critical point, we observe a learning DP (LDP) behavior, where some observables follow the DP scaling, while others show a distinct behavior affected by learning. 
At the spreading critical point, in higher dimensions, our results suggest a critical point in the universality class of dynamical percolation (DyP), as in the SIR model. In this case, learning with negative reinforcement turns the SIS model (without immunization) into an effective SIR model, even though the local infection and healing rates remain unaffected.
In the region between the spreading and the bulk critical point, the system is in a phase showcasing a propagating front with an subcritical bulk. It is an interesting future direction to explore if such a phase could be observed in some real-life applications, either in infection, brain activity or population spreading.  
%
Although the contact process is a genuinely non-equilibrium model and learning leads to changes in the model parameters coupled to the dynamics, we were also able to obtain informative analytic results, further supporting our simulation observations. 

In the target learning model, brain-inspired positive reinforcement leads to a Griffiths phase in higher dimensions. There is increasing evidence that brain dynamics shows hallmarks of being in a Griffiths phase \cite{buendia2022broad, Girardi-Schappo2016, PhysRevResearch.6.023018, Fusca2023}. 
However, in standard models, a Griffiths phase is either caused by uncorrelated time-invariant (quenched) disorder \cite{IgloiMonthus2005}, or a hierarchical-modular architecture of cortical networks \cite{MorettiMunoz2013, Odor2015}. Our results serve as an independent alternative mechanism, a manifestation of the ant mill phenomenon \cite{Erhard2022}, originating from dynamical trapping in activity loops emerging from correlated heterogeneities induced by learning. 
In the brain, it is believed that being in a Griffiths phase offers functional advantages, in terms of memory and computational capabilities \cite{MorettiMunoz2013}.
Signatures of Griffiths phases might be also present in artifical neural networks \cite{Logan}, an exciting future direction to explore in relation to Hebbian learning.

The target learning model also leads to an emergent quenched system in the $r\to0$ limit, leading to invariant directional configuration(s) in both 1D (totally asymmetric CP \cite{tretyakov}) and 2D. Note that in 2D the system is disordered, as there are infinitely many directional configurations depending on the initial condition and the dynamics. The directional configuration also contains correlations resulting from the generating dynamics. Without learning, uncorrelated quenched disorder in the rates leads to a distinct type of critical behavior characterized by ultra-slow activated scaling \cite{IgloiMonthus2005} below four dimensions \cite{VojtaFarquharMast2009, Vojta2012} and on certain types of complex networks \cite{Juhasz_2013}, in line with the perturbative argument known as the Harris criterion \cite{Harris1974criterion}. It is an interesting open question to explore whether such a directional quenched disorder is a relevant perturbation at the critical point, potentially leading to exotic, strong disorder phenomena.

To the best of our knowledge, our work presented here is the first study of Hebbian learning for activity spreading, with several future directions remaining open. Such extensions could consider alternative forms of Hebbian learning, either incorporating information on the duration when the sites are simultaneously active, or different normalization schemes. For example, in the case of positive reinforcement, brain dynamics would suggest the study of the potentially stabilizing impact of inherently decaying activation rates \cite{Miller}.
As real-life infection spreading is complicated, combinations with other mechanisms, like the simultaneous impact of Hebbian learning and awareness \cite{Odor2025, PhysRevResearch.7.L012061} could lead to more realistic models.

Although the CP is equivalent to the SIS model on a lattice, this is no longer true on complex networks. In stark contrast to the CP \cite{PhysRevLett.96.038701}, in the SIS model there is no inactive phase on scale-free networks with degree exponent $\gamma>2$ \cite{PhysRevE.63.066117, PhysRevLett.105.218701, Durrett}. Hence, positive reinforcement might not completely eliminate the active phase. Instead, a phase transition might happen between an inactive  and an active phase on such networks. 
Another direction of study is unveiling the condition under which a similar reinforcement-driven Griffiths phase occurs on complex networks through the ant mill phenomenon \cite{Erhard2022}.

\begin{acknowledgments}
This work was supported by the National Research, Development and Innovation Office NKFIH under Grant No.~K146736. The work of I.A.K. was supported by the National Science Foundation under Grant No.~PHY-2310706 of the QIS program in the Division of Physics, by the Domus Hungary Scholarship of the Hungarian Academy of Sciences, and by the Baker Faculty Grant of the Weinberg College of Arts and Sciences, Northwestern University. W.T.E. was supported by the Northwestern Office of Undergraduate Research's Summer Undergraduate Research Grant (SURG) and the Baker Program in Undergraduate Research. We thank L.~V.~Luzzatto, N.~Love, and J.~Barrera Lopez for useful discussions, and M.~M.~Driscoll, B.~Oborny and G.~\'Odor for useful comments.
\end{acknowledgments}


\appendix
\section{Imbalance in one dimension}
\label{sec:analytic}

\subsection{Standard contact process}
\label{subsec:cp}

For the standard CP of $r=1$ the rates are constant, i.e.~independent from the imbalance variables $n_i$.
According to our numerical observations, the distribution $p_n$ gets broader algebraically as $t^{\kappa}$. In the active phase, the correlation time $\tau$ is finite. Therefore, the imbalance variables $n_i(t_m)$ sampled at times $t_m=mT$, where $T\gg \tau$ and $m=0,1,2,\dots$, are expected to have uncorrelated increments, leading to $\kappa=1/2$. This is indeed in agreement with our simulations (not shown). At the critical point, however, we have found that the scaling of the distribution $p_n$ with time is characterized by a smaller exponent $\kappa=0.43(1)$, see Fig.~\ref{fig_ndist}a.   
Besides possible temporal correlations in the process $n_i(t)$, there is also an obvious reason for the difference between the exponent $\kappa$ in the active phase and at the critical point. Clearly, the change of imbalance variables is triggered by local activity and the 'clock' of this process is ticking only if the corresponding site is active, otherwise it is still. In the active phase, the mean local activity is constant in time in the steady state. As opposed to this, at the critical point, it decreases as $\rho\sim t^{-\delta}$. As a consequence, the time $t'$ relevant for the process of imbalance variables at the critical point is related to the real time $t$ as $t'\sim \int^t\rho(t^{''})dt^{''}\sim t^{1-\delta}$.
Ignoring temporal correlations, the width of the distribution would then increase proportionally to $t'^{1/2}\sim t^{(1-\delta)/2}$, yielding 
\be 
\kappa=\frac{1-\delta}{2}=0.4203 
\ee
at the critical point. This is compatible with the numerical estimate, suggesting that temporal correlations play a weak role, if any. 

\subsection{Source learning model}

\subsubsection{Dynamics}

The imbalance $n$ keeps track of the cumulative number of successful activation events from a given site, which counts the activations to the right (left)  with a positive (negative) sign. The activation rate to the right is $r_n=\lambda\frac{r^n}{1+r^n}$, whereas the rate to the left $l_n$ is fixed by $l_n+r_n=\lambda$. Thus, $r_n=l_{-n}$. In the source learning model, the imbalance of the source site is increased (decreased) by $1$ after each successful activation to the right (left). 
We will denote the state of an active (inactive) site with imbalance $n$ by $A_n$ ($I_n$). The probability of finding a site in state $A_n$ and $I_n$ is given by $p_n$ and $q_n$, respectively. The probability that the site is active is $p\equiv\sum_np_n$, and we also introduce $q\equiv 1-p$.

Let us consider a mean-field approximation of model, in which the local states of different sites are assumed to be uncorrelated.
For a technical convenience, we will restrict ourselves in the subsequent analysis to the domain $r>0$. 
The terms contributing to the change rate of $p_n$ at a marked site  are then the following. i) The loss term by spontaneous deactivations: $-\mu p_n$.
ii) The loss by activation events from the marked site to the right, $-r_np_nq$, and to the left, $-l_np_nq$. In total $-\lambda p_nq$.
iii) The gain by activation events from the marked site to the right $r_{n-1}p_{n-1}q$, and to the left, $l_{n+1}p_{n+1}q$.
iv) The gain by the activation of the marked site from the left, $(\sum_mr_mp_m)q_n$, and from the right,  $(\sum_ml_mp_m)q_n$. In total $\lambda pq_n$.
The rate of change of $p_n$ is then
\be
\dot{p}_n = -\mu p_n -\lambda p_nq + r_{n-1}p_{n-1}q + l_{n+1}p_{n+1}q + \lambda pq_n.
\label{pn}
\ee
Summing over $n$, we obtain
\be
\dot{p}=-\mu p + \lambda p(1-p),
\label{mfcp}
\ee
which coincides with the mean-field description of the standard CP. Thus, within this approximation, the reinforcement does not shift the location of the critical point, $\lambda_c/\mu=1$.

Now we turn to the rate of change of $q_n$. The corresponding terms are the following.
i) There is gain by spontaneous deactivations, $\mu p_n$.
ii) There is a loss term by activations of the marked site from the left, $-(\sum_mr_mp_m)q_n$ and from the right, $-(\sum_ml_mp_m)q_n$. In total $-\lambda pq_n$.
Hence, we have
\be
\dot{q}_n= \mu p_n - \lambda pq_n.
\label{qn}
\ee

\begin{figure*}[ht!]
\includegraphics[width=0.48\linewidth]{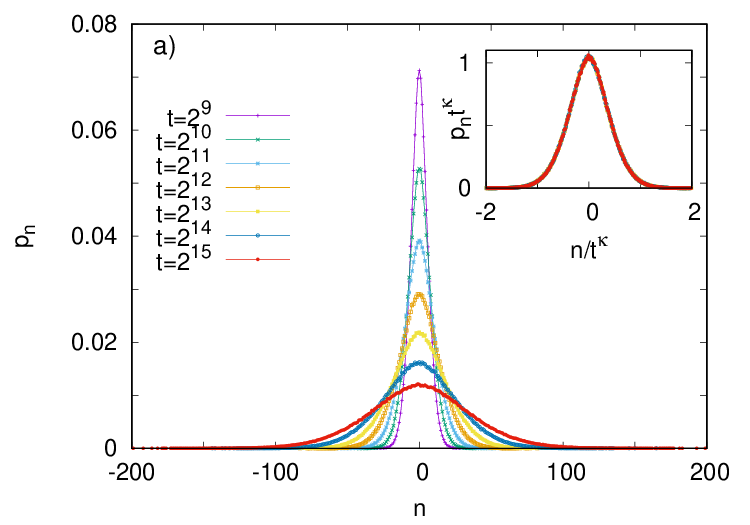}%
\includegraphics[width=0.48\linewidth]{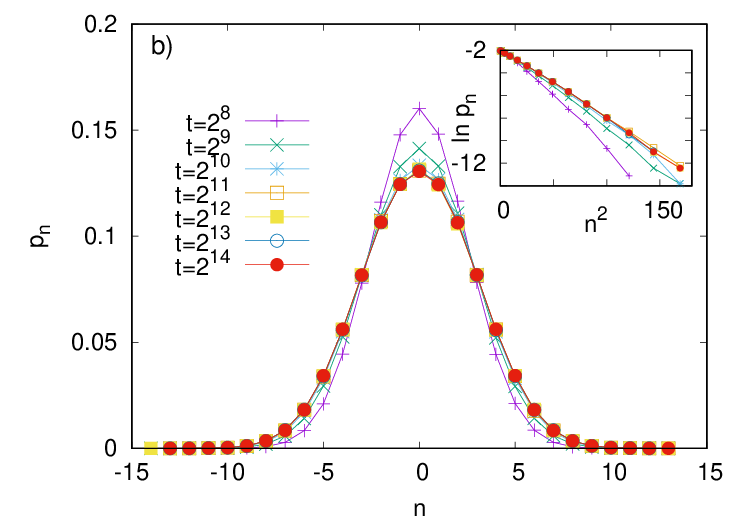}%
\vskip-3mm
\caption{\justifying \textbf{Imbalance in the source learning model.} a) Distribution of the imbalance variable $n_i$ at the critical point of the model with $r=1$, measured at different times starting from a fully active initial state. The inset shows the scaled distributions, where a scaling collapse is obtained by using 
$\kappa=0.43$. 
b) Distribution of the imbalance variable $n_i$ in the active phase of the source learning model with $r=0.8$ ($\lambda=4$). The inset demonstrates the Gaussian tail of the limit distribution, corresponding to a straight line with a slope $C=0.064(2)$. 
}
\label{fig_ndist}
\end{figure*}

\subsubsection{Stationary state}

In a weak sense, $\dot{p}=0$, the model has a non-trivial ($p>0$) stationary state above the threshold, $\lambda/\mu>1$, for any $r>0$.
However, non-trivial stationarity in a stronger sense, $\dot{p}_n=\dot{q}_n=0$, exists only for $0<r<1$ i.e. for negative reinforcement. 

To see this, let us assume that the process is stationary in the weak sense. Then Eq.~(\ref{mfcp}) implies that the probability of finding a state in an active state is constant, $p=1-\mu/\lambda$. 
In this case, Eqs.~(\ref{pn}) and (\ref{qn}) can be regarded as the master equation of a continuous-time random walk on a comb-like lattice. The allowed transitions are nearest-neighbor hops in the set of active states, $A_n\to A_{n+1}$ and $A_n\to A_{n-1}$, with rates $r_nq$ and $l_nq$, respectively. As well as hops to ``dangling" states $I_n$ back and forth, $A_n\to I_n$ and $I_n\to A_n$, with rates $\mu$ and $\lambda p$, respectively. 
It is easy to see that this random walk will be recurrent only if $0<r<1$, in which case it has a stationary state in the strong sense, $\dot{p}_n=\dot{q}_n=0$.
This leads to
\be
q_n=\frac{\mu}{\lambda}\frac{p_n}{p}
\label{qnstac}
\ee
and
\be
0=-\lambda p_n + r_{n-1}p_{n-1} + l_{n+1}p_{n+1}.
\ee
As the transition graph is acyclic, detailed balance holds in the stationary state, $r_np_n=l_{n+1}p_{n+1}$. The stationary probabilities can then be written in terms of a potential $U_n$, as $p_n=\mathcal{N}e^{-U_n}$, where the potential differences are fixed by $U_{n+1}-U_{n}=\ln\frac{l_{n+1}}{r_n}$ and $\mathcal{N}$ denotes a normalization constant. 
Let us now consider the large-$n$ asymptotics of stationary probabilities $p_n$ in the case $0<r<1$.
For $n\gg 1$, we have $r_n\simeq\lambda r^n$ and $l_n\simeq \lambda$. The potential difference is then $U_{n+1}-U_{n}\simeq n\ln(1/r)$. The potential for large $n$ is thus $U_n\simeq n^2\frac{\ln(1/r)}{2}$.
The stationary probability distribution of the imbalance has therefore the tail
\be
p_n\sim e^{-U_n}\simeq e^{-n^2\frac{\ln(1/r)}{2}}.
\ee
This Gaussian tail of the imbalance distribution is also observed in simulations, see Fig.~\ref{fig_ndist}. 
Note that the characteristic width of the imbalance distribution is $n^*\sim [\ln(1/r)]^{-1/2}$, which is diverging as $r\to 1$. Nevertheless, the asymmetry between activation rates to the right and to the left, which can be conveniently measured by $a_n\equiv \ln(r_n/l_n)$, has a characteristic value $a^*\sim \sqrt{\ln(1/r)}$, which is vanishing as $r\to 1$.

In the case of a non-negative reinforcement, $r\ge 1$, the random walk is no longer stable, and there exists no stationary state. At $r=1$, the walker spreads out diffusively, $\langle n^2\rangle\sim t$, while for $r>1$, the spreading  is ballistic, $\langle n^2\rangle\sim t^2$.
For a weak positive reinforcement, $r\gtrsim 1$, the dynamics of the imbalance variable will display a crossover phenomenon from the non-reinforced behavior at short times, where the random walk is recurrent to the positively reinforced case at late times, where the random walk is transient. The crossover time scale $\tau$ will diverge as $r\to 1+$. The form of the divergence within the mean-field approximation can be obtained as follows. 
The potential for $n\ge 1$ can be evaluated to be $U_n-U_0=-\frac{n(n+1)}{2}\ln r -\ln (\frac{1+r^n}{2})$.  
If $r$ is close to $1$, we have $U_n-U_0=-\frac{n^2}{2}(r-1) + O(n^2(r-1)^2)$. 
The crossover from recurrence to transience occurs when the walk reaches a point of no return $n^*$ which is marked by the condition $U_{n^*}-U_0\sim O(1)$. This scales as $n^*\sim (r-1)^{-1/2}$, and, since the random walk is essentially unbiased for $|n|\ll n^*$, we find for the crossover time scale 
\be 
\tau\sim (n^*)^2\sim \frac{1}{r-1}
\label{divtau}
\ee
as $r\to 1+$, in line with our numerical results in Fig.~\ref{fig_tau_d}.

\subsection{Mutual learning model}

\subsubsection{Dynamics}

The mutual learning model differs from the source learning model in that, after a successful activation event, also the imbalance of the target site is changed by $1$, with an opposite sign compared to that at the source site.
Concerning the change rate of $p_n$, this extension modifies only term iv) in the previous section. Here, the gain by activation events of the marked site from the left will be  $(\sum_mr_mp_m)q_{n+1}$, while from the right  $(\sum_ml_mp_m)q_{n-1}$.
The rate of change of $p_n$ is thus governed by
\beqn
\dot{p}_n = -\mu p_n -\lambda p_nq + r_{n-1}p_{n-1}q + l_{n+1}p_{n+1}q +
\nonumber \\
+ \overline{rp}pq_{n+1} + \overline{lp}pq_{n-1},
\label{pnu}
\eeqn
where  $\overline{rp}\equiv\frac{1}{p}\sum_mr_mp_m$ and $\overline{lp}\equiv\frac{1}{p}\sum_ml_mp_m$.
Summing over $n$ leads again to Eq.~(\ref{mfcp}), thus $\lambda_c/\mu=1$ holds also in the mean-field approximation of the mutual learning model.
The rate of change of $q_n$ is the same as in the source learning model, see Eq.~(\ref{qn}). 

\begin{figure}[ht!]
\includegraphics[width=0.96\linewidth]{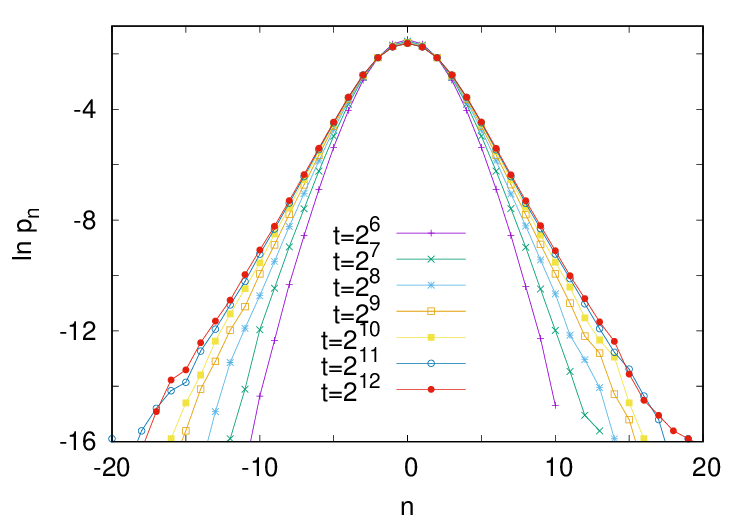}%
\vskip-3mm
\caption{\justifying \textbf{Imbalance in the mutual learning model.} Distribution of the variable $n_i$ in the active phase ($\lambda=4$) of the mutual learning model with $r=0.5$, measured at different times starting from a fully active initial state. We observe an exponential tail $p_n\sim e^{-an}$, with $a=0.87(2)$. 
}
\label{fig_ndist_un}
\end{figure}

\subsubsection{Stationary state}

Similarly to the source learning model, there is a non-trivial stationary state above the threshold in the weak sense for any $r>0$.
In the stronger sense, a non-trivial stationary state exists again only for $0<r<1$.

Let us again assume weak stationarity, and for the sake of simplicity, let us also assume that the initial state is symmetric under $n\to -n$. Then it will remain so at any time, and, as a consequence $\overline{lp}=\overline{rp}=\lambda/2$, i.e.  $\overline{lp}$  and $\overline{rp}$ are constant in time.
Eqs.~(\ref{pnu}) and (\ref{qn}) can then be interpreted as a master equation of a random walk on a ladder with the following transitions. $A_n\to A_{n-1}$ and  $A_n\to A_{n+1}$ with rates $l_nq$ and $r_nq$, respectively. Furthermore, hopping to the other leg of the ladder, $A_n\to I_n$, with rate $\mu$, and hopping back diagonally as $I_n\to A_{n\pm 1}$ with rate $\frac{\lambda}{2}p$. 
As in the source learning case, this random walk will be recurrent and have a stationary state only if $0<r<1$.
Setting $\dot{p}_n=\dot{q}_n=0$ in the dynamical equations, we obtain Eq.~(\ref{qnstac}) and
\be
0=-2\lambda p_n + \left(r_{n-1}+\frac{\lambda}{2}\right)p_{n-1} + \left(l_{n+1}+\frac{\lambda}{2}\right)p_{n+1}.
\ee
Thus, the stationary probabilities $p_n$ are, up to a normalization constant, the stationary probabilities of a random walk on the integers with hopping rates $\omega_{n,n+1}=r_n+\frac{\lambda}{2}$ and $\omega_{n,n-1}=l_n+\frac{\lambda}{2}$.
Let us consider the asymptotics of the imbalance distribution for $0<r<1$. The potential difference of the corresponding random walk for large $n$ is given by 
$U_{n+1}-U_{n}=\ln\frac{\omega_{n+1,n}}{\omega_{n,n+1}}\simeq \ln 3$.
Thus, $U_n\simeq n\ln 3$, and the tail of the stationary imbalance distribution is exponential:
\be
p_n\sim e^{-U_n}\simeq e^{-n\ln 3}.
\ee
As discussed in the main text, this functional form is again in agreement with numerical simulations, although with a modified decay rate, see Fig.~\ref{fig_ndist_un}. 

\begin{figure*}[ht!]
\includegraphics[width=0.4\linewidth]{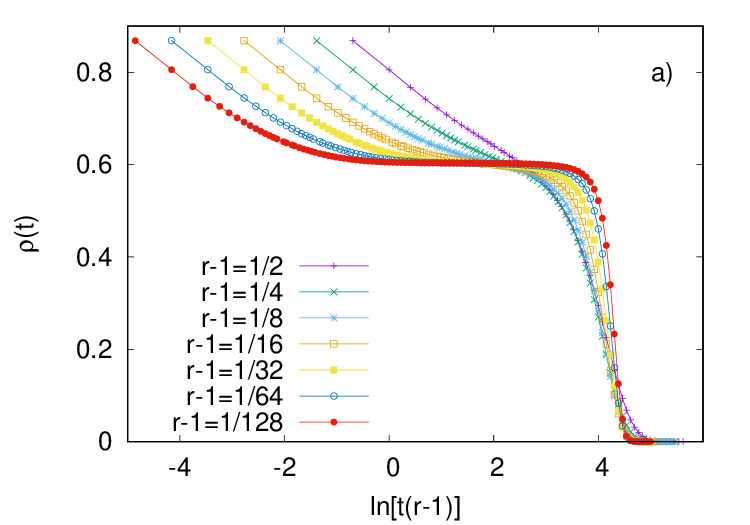}%
\includegraphics[width=0.4\linewidth]{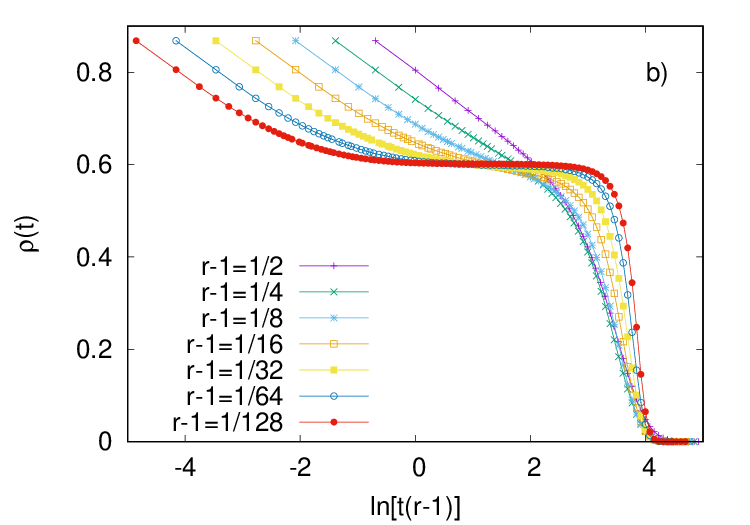}%
\\
\includegraphics[width=0.4\linewidth]{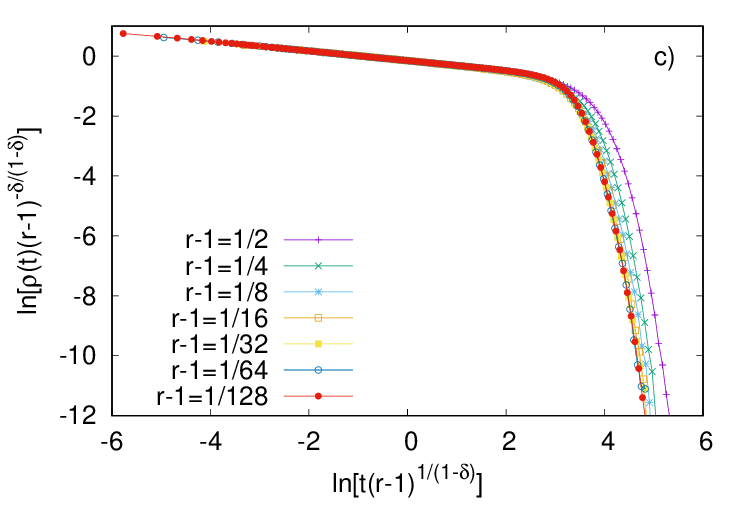}%
\includegraphics[width=0.4\linewidth]{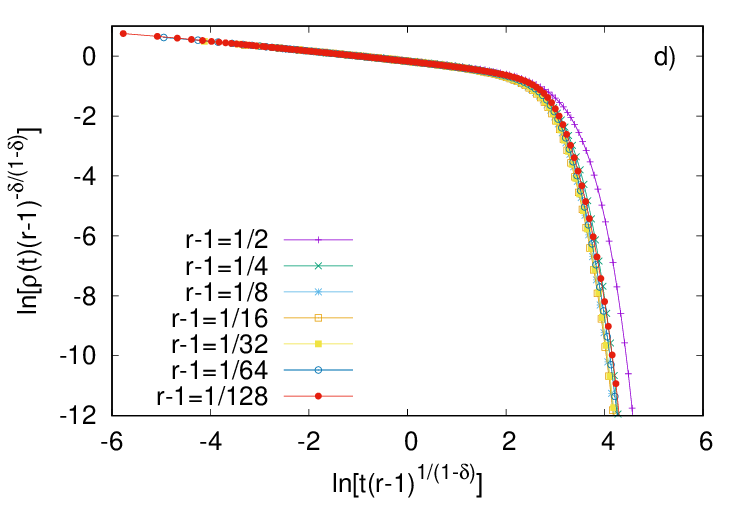}%
\vskip-3mm
\caption{\justifying \textbf{Scaling of lifetime for weak positive reinforcement in one dimension.} a) \& b) Time dependence of global density in the active phase ($\lambda=4$) for different values of the reinforcement parameter in the source (a) and mutual (b) learning models. Time is rescaled as $t\to \tilde t=t(r-1)$. 
 c) \& d)  The same at the critical activation rate of the standard CP ($r=1$) in the source (c) and mutual (d) learning models. Time is rescaled as $t\to \tilde t=t(r-1)^{1/(1-\delta)}$.
}
\label{fig_tau_d}
\end{figure*}

In the case of non-negative reinforcement, $r\ge 1$, the random walk process is unstable and $\langle n^2\rangle$ increases unboundedly, in the same way as in the source learning model. 
The divergence of the time scale as $r\to 1+$, can be obtained in the same way as for the source learning model. Here, we one has to consider that although the potential varies linearly for $|n|\gg (r-1)^{-1}$, the point of no return is closer to the origin, scaling as $n^*\sim (r-1)^{-1/2}$. Indeed, for small $r-1$, the potential is parabolic around the origin,  
 $U_n-U_0=-\frac{n^2}{4}(r-1) + O((r-1)^2n^3)$. This leads to the same form of the divergence of time scale as found in Eq.~(\ref{divtau}) for the mutual learning model, see Fig.~\ref{fig_tau_d}.

\subsection{Target learning model}

In the target learning model, after a successful activation event, the imbalance of the target site is updated (in the same way as for the mutual learning model) but the imbalance of the source site is left unchanged. 
Within a mean-field approximation which ignores correlations between occupation and imbalance variables at different sites, the rate of change of $p_n$ reads in one dimension as 
\be
\dot{p}_n = -\mu p_n + \overline{rp}pq_{n+1} + \overline{lp}pq_{n-1},
\label{pnt}
\ee
while that of $q_n$ is the same as given in Eq.~(\ref{qn}).
Assuming weak stationarity and symmetry under $n\to -n$, these equations describe a random walk on a ladder with vertical transitions $A_n\to I_n$ with rate $\mu$, and diagonal jumps $I_n\to A_{n\pm 1}$ with rate $\lambda p/2$.
This leads to an unbounded diffusive broadening of the imbalance variable as  $\langle n^2\rangle \sim t$, irrespective of the value of the reinforcement parameter $r$. Therefore, there is no stationary state in this case.

\section{Positive reinforcement in one dimension}
\label{sec:pos}

For positive reinforcement with $r>1$ there is no active phase, as the active sites form isolated localized clusters that eventually inactivate without triggering a macroscopic active phase. 
The initial state $n_i=0$ becomes unstable and the local imbalance variables start to drift either in the positive or in the negative direction, which is singled out randomly by stochastic fluctuations. The return to the origin ($n_i=0$) becomes more and more unlikely with increasing $|n_i|$, thus the symmetry of local variables present for $r<1$ is spontaneously broken here. 
Evidently, an active stationary state would then lead to extreme random imbalance $|n_i|\to\infty$, which, however, cannot sustain an active steady state. As a consequence, a non-trivial steady state with a positive density cannot exist in the case $r>1$, which is in agreement with our numerical observations.  
The state of the system when started from a fully active state with zero imbalance will therefore tend to the absorbing state at late times. This process occurs on a finite time scale, the characteristic lifetime $\tau$, which depends on $r$ and expected to diverge as $r\to 1+$. According to mean-field theory presented in the Appendix~\ref{sec:analytic}, the divergence is of the form $\tau\sim 1/(r-1)$ above the critical activation rate of the standard CP ($r=1$). 
Numerical results presented in Fig.~\ref{fig_tau_d} indeed show an abrupt decline in the time dependence of density at a characteristic time diverging in accordance with the prediction of mean-field theory. 
At the critical activation rate of the standard CP, $\lambda=\lambda_c$, this is to be corrected by the replacement $\tau\to \tau^{1-\delta}$, yielding $\tau\sim (r-1)^{-1/(1-\delta)}$, see Sec. \ref{subsec:cp}. Numerical results shown in Fig.~\ref{fig_tau_d} are agreement with this prediction.
For mutual learning the only difference is that at $\lambda=\lambda_c$, we find a slight deviation from the expectation $\tau\sim (r-1)^{-1/(1-\delta)}$, see Fig.~\ref{fig_tau_d}. The optimal collapse (not shown) is obtained by an exponent $1.25$ instead of $1/(1-\delta)=1.1897$.  

\bibliography{paperrefs}

\begin{thebibliography}{57}%
\makeatletter
\providecommand \@ifxundefined [1]{%
 \@ifx{#1\undefined}
}%
\providecommand \@ifnum [1]{%
 \ifnum #1\expandafter \@firstoftwo
 \else \expandafter \@secondoftwo
 \fi
}%
\providecommand \@ifx [1]{%
 \ifx #1\expandafter \@firstoftwo
 \else \expandafter \@secondoftwo
 \fi
}%
\providecommand \natexlab [1]{#1}%
\providecommand \enquote  [1]{``#1''}%
\providecommand \bibnamefont  [1]{#1}%
\providecommand \bibfnamefont [1]{#1}%
\providecommand \citenamefont [1]{#1}%
\providecommand \href@noop [0]{\@secondoftwo}%
\providecommand \href [0]{\begingroup \@sanitize@url \@href}%
\providecommand \@href[1]{\@@startlink{#1}\@@href}%
\providecommand \@@href[1]{\endgroup#1\@@endlink}%
\providecommand \@sanitize@url [0]{\catcode `\\12\catcode `\$12\catcode `\&12\catcode `\#12\catcode `\^12\catcode `\_12\catcode `\%12\relax}%
\providecommand \@@startlink[1]{}%
\providecommand \@@endlink[0]{}%
\providecommand \url  [0]{\begingroup\@sanitize@url \@url }%
\providecommand \@url [1]{\endgroup\@href {#1}{\urlprefix }}%
\providecommand \urlprefix  [0]{URL }%
\providecommand \Eprint [0]{\href }%
\providecommand \doibase [0]{https://doi.org/}%
\providecommand \selectlanguage [0]{\@gobble}%
\providecommand \bibinfo  [0]{\@secondoftwo}%
\providecommand \bibfield  [0]{\@secondoftwo}%
\providecommand \translation [1]{[#1]}%
\providecommand \BibitemOpen [0]{}%
\providecommand \bibitemStop [0]{}%
\providecommand \bibitemNoStop [0]{.\EOS\space}%
\providecommand \EOS [0]{\spacefactor3000\relax}%
\providecommand \BibitemShut  [1]{\csname bibitem#1\endcsname}%
\let\auto@bib@innerbib\@empty
\bibitem [{\citenamefont {Rosen}(1975)}]{ROSEN197539}%
  \BibitemOpen
  \bibfield  {author} {\bibinfo {author} {\bibfnamefont {R.}~\bibnamefont {Rosen}},\ }\bibfield  {title} {\bibinfo {title} {Biological systems as paradigms for adaptation},\ }in\ \href {https://doi.org/https://doi.org/10.1016/B978-0-12-207350-2.50005-1} {\emph {\bibinfo {booktitle} {Adaptive Economic Models}}},\ \bibinfo {editor} {edited by\ \bibinfo {editor} {\bibfnamefont {R.~H.}\ \bibnamefont {Day}}\ and\ \bibinfo {editor} {\bibfnamefont {T.}~\bibnamefont {Groves}}}\ (\bibinfo  {publisher} {Academic Press},\ \bibinfo {year} {1975})\ pp.\ \bibinfo {pages} {39--72}\BibitemShut {NoStop}%
\bibitem [{\citenamefont {Guirfa}\ and\ \citenamefont {Menzel}(2003)}]{Guirfa2003}%
  \BibitemOpen
  \bibfield  {author} {\bibinfo {author} {\bibfnamefont {M.}~\bibnamefont {Guirfa}}\ and\ \bibinfo {author} {\bibfnamefont {R.}~\bibnamefont {Menzel}},\ }\bibinfo {title} {Biology of adaptation and learning},\ in\ \href {https://doi.org/10.1007/978-3-662-05594-6_2} {\emph {\bibinfo {booktitle} {Adaptivity and Learning: An Interdisciplinary Debate}}},\ \bibinfo {editor} {edited by\ \bibinfo {editor} {\bibfnamefont {R.}~\bibnamefont {K{\"u}hn}}, \bibinfo {editor} {\bibfnamefont {R.}~\bibnamefont {Menzel}}, \bibinfo {editor} {\bibfnamefont {W.}~\bibnamefont {Menzel}}, \bibinfo {editor} {\bibfnamefont {U.}~\bibnamefont {Ratsch}}, \bibinfo {editor} {\bibfnamefont {M.~M.}\ \bibnamefont {Richter}},\ and\ \bibinfo {editor} {\bibfnamefont {I.-O.}\ \bibnamefont {Stamatescu}}}\ (\bibinfo  {publisher} {Springer Berlin Heidelberg},\ \bibinfo {address} {Berlin, Heidelberg},\ \bibinfo {year} {2003})\ pp.\ \bibinfo {pages} {7--9}\BibitemShut {NoStop}%
\bibitem [{\citenamefont {Tu}\ and\ \citenamefont {Rappel}(2018)}]{turev}%
  \BibitemOpen
  \bibfield  {author} {\bibinfo {author} {\bibfnamefont {Y.}~\bibnamefont {Tu}}\ and\ \bibinfo {author} {\bibfnamefont {W.-J.}\ \bibnamefont {Rappel}},\ }\bibfield  {title} {\bibinfo {title} {Adaptation in living systems},\ }\href {https://doi.org/https://doi.org/10.1146/annurev-conmatphys-033117-054046} {\bibfield  {journal} {\bibinfo  {journal} {Annual Review of Condensed Matter Physics}\ }\textbf {\bibinfo {volume} {9}},\ \bibinfo {pages} {183} (\bibinfo {year} {2018})}\BibitemShut {NoStop}%
\bibitem [{\citenamefont {Harris}(1974{\natexlab{a}})}]{harris1974}%
  \BibitemOpen
  \bibfield  {author} {\bibinfo {author} {\bibfnamefont {T.~E.}\ \bibnamefont {Harris}},\ }\bibfield  {title} {\bibinfo {title} {Contact interactions on a lattice},\ }\href@noop {} {\bibfield  {journal} {\bibinfo  {journal} {The Annals of Probability}\ }\textbf {\bibinfo {volume} {2}},\ \bibinfo {pages} {969} (\bibinfo {year} {1974}{\natexlab{a}})}\BibitemShut {NoStop}%
\bibitem [{\citenamefont {Hebb}(2005)}]{Hebb2005}%
  \BibitemOpen
  \bibfield  {author} {\bibinfo {author} {\bibfnamefont {D.~O.}\ \bibnamefont {Hebb}},\ }\href@noop {} {\emph {\bibinfo {title} {The organization of behavior: A neuropsychological theory}}}\ (\bibinfo  {publisher} {Psychology press},\ \bibinfo {year} {2005})\BibitemShut {NoStop}%
\bibitem [{\citenamefont {Magnasco}\ \emph {et~al.}(2009)\citenamefont {Magnasco}, \citenamefont {Piro},\ and\ \citenamefont {Cecchi}}]{Magnasco2009-we}%
  \BibitemOpen
  \bibfield  {author} {\bibinfo {author} {\bibfnamefont {M.~O.}\ \bibnamefont {Magnasco}}, \bibinfo {author} {\bibfnamefont {O.}~\bibnamefont {Piro}},\ and\ \bibinfo {author} {\bibfnamefont {G.~A.}\ \bibnamefont {Cecchi}},\ }\bibfield  {title} {\bibinfo {title} {Self-tuned critical anti-{H}ebbian networks},\ }\href@noop {} {\bibfield  {journal} {\bibinfo  {journal} {Phys. Rev. Lett.}\ }\textbf {\bibinfo {volume} {102}},\ \bibinfo {pages} {258102} (\bibinfo {year} {2009})}\BibitemShut {NoStop}%
\bibitem [{\citenamefont {te~Vrugt}\ \emph {et~al.}(2020)\citenamefont {te~Vrugt}, \citenamefont {Bickmann},\ and\ \citenamefont {Wittkowski}}]{teVrugt2020}%
  \BibitemOpen
  \bibfield  {author} {\bibinfo {author} {\bibfnamefont {M.}~\bibnamefont {te~Vrugt}}, \bibinfo {author} {\bibfnamefont {J.}~\bibnamefont {Bickmann}},\ and\ \bibinfo {author} {\bibfnamefont {R.}~\bibnamefont {Wittkowski}},\ }\bibfield  {title} {\bibinfo {title} {Effects of social distancing and isolation on epidemic spreading modeled via dynamical density functional theory},\ }\href {https://doi.org/10.1038/s41467-020-19024-0} {\bibfield  {journal} {\bibinfo  {journal} {Nature Communications}\ }\textbf {\bibinfo {volume} {11}},\ \bibinfo {pages} {5576} (\bibinfo {year} {2020})}\BibitemShut {NoStop}%
\bibitem [{\citenamefont {Moretti}\ and\ \citenamefont {Muñoz}(2013)}]{MorettiMunoz2013}%
  \BibitemOpen
  \bibfield  {author} {\bibinfo {author} {\bibfnamefont {P.}~\bibnamefont {Moretti}}\ and\ \bibinfo {author} {\bibfnamefont {M.~A.}\ \bibnamefont {Muñoz}},\ }\bibfield  {title} {\bibinfo {title} {Griffiths phases and the stretching of criticality in brain networks},\ }\href {https://doi.org/10.1038/ncomms3521} {\bibfield  {journal} {\bibinfo  {journal} {Nature Communications}\ }\textbf {\bibinfo {volume} {4}},\ \bibinfo {pages} {2521} (\bibinfo {year} {2013})}\BibitemShut {NoStop}%
\bibitem [{\citenamefont {Moldakarimov}\ and\ \citenamefont {Sejnowski}(2008)}]{MOLDAKARIMOV2008667}%
  \BibitemOpen
  \bibfield  {author} {\bibinfo {author} {\bibfnamefont {S.}~\bibnamefont {Moldakarimov}}\ and\ \bibinfo {author} {\bibfnamefont {T.}~\bibnamefont {Sejnowski}},\ }\bibfield  {title} {\bibinfo {title} {1.34 - neural computation theories of learning},\ }in\ \href {https://doi.org/https://doi.org/10.1016/B978-012370509-9.00086-3} {\emph {\bibinfo {booktitle} {Learning and Memory: A Comprehensive Reference}}},\ \bibinfo {editor} {edited by\ \bibinfo {editor} {\bibfnamefont {J.~H.}\ \bibnamefont {Byrne}}}\ (\bibinfo  {publisher} {Academic Press},\ \bibinfo {address} {Oxford},\ \bibinfo {year} {2008})\ pp.\ \bibinfo {pages} {667--679}\BibitemShut {NoStop}%
\bibitem [{\citenamefont {Clark}\ and\ \citenamefont {Abbott}(2024)}]{PhysRevX.14.021001}%
  \BibitemOpen
  \bibfield  {author} {\bibinfo {author} {\bibfnamefont {D.~G.}\ \bibnamefont {Clark}}\ and\ \bibinfo {author} {\bibfnamefont {L.~F.}\ \bibnamefont {Abbott}},\ }\bibfield  {title} {\bibinfo {title} {Theory of coupled neuronal-synaptic dynamics},\ }\href {https://doi.org/10.1103/PhysRevX.14.021001} {\bibfield  {journal} {\bibinfo  {journal} {Phys. Rev. X}\ }\textbf {\bibinfo {volume} {14}},\ \bibinfo {pages} {021001} (\bibinfo {year} {2024})}\BibitemShut {NoStop}%
\bibitem [{\citenamefont {Lynn}\ \emph {et~al.}(2024)\citenamefont {Lynn}, \citenamefont {Holmes},\ and\ \citenamefont {Palmer}}]{Lynn2024}%
  \BibitemOpen
  \bibfield  {author} {\bibinfo {author} {\bibfnamefont {C.~W.}\ \bibnamefont {Lynn}}, \bibinfo {author} {\bibfnamefont {C.~M.}\ \bibnamefont {Holmes}},\ and\ \bibinfo {author} {\bibfnamefont {S.~E.}\ \bibnamefont {Palmer}},\ }\bibfield  {title} {\bibinfo {title} {Heavy-tailed neuronal connectivity arises from {H}ebbian self-organization},\ }\href {https://doi.org/10.1038/s41567-023-02332-9} {\bibfield  {journal} {\bibinfo  {journal} {Nature Physics}\ }\textbf {\bibinfo {volume} {20}},\ \bibinfo {pages} {484} (\bibinfo {year} {2024})}\BibitemShut {NoStop}%
\bibitem [{\citenamefont {Oborny}\ \emph {et~al.}(2007)\citenamefont {Oborny}, \citenamefont {Szabó},\ and\ \citenamefont {Meszéna}}]{Oborny}%
  \BibitemOpen
  \bibfield  {author} {\bibinfo {author} {\bibfnamefont {B.}~\bibnamefont {Oborny}}, \bibinfo {author} {\bibfnamefont {G.}~\bibnamefont {Szabó}},\ and\ \bibinfo {author} {\bibfnamefont {G.}~\bibnamefont {Meszéna}},\ }\bibinfo {title} {Survival of species in patchy landscapes: percolation in space and time},\ in\ \href@noop {} {\emph {\bibinfo {booktitle} {Scaling Biodiversity}}},\ \bibinfo {series and number} {Ecological Reviews},\ \bibinfo {editor} {edited by\ \bibinfo {editor} {\bibfnamefont {D.}~\bibnamefont {Storch}}, \bibinfo {editor} {\bibfnamefont {P.}~\bibnamefont {Marquet}},\ and\ \bibinfo {editor} {\bibfnamefont {J.}~\bibnamefont {Brown}}}\ (\bibinfo  {publisher} {Cambridge University Press},\ \bibinfo {year} {2007})\ p.\ \bibinfo {pages} {409–440}\BibitemShut {NoStop}%
\bibitem [{\citenamefont {Best}\ \emph {et~al.}(2010)\citenamefont {Best}, \citenamefont {White}, \citenamefont {Kisdi}, \citenamefont {Antonovics}, \citenamefont {Brockhurst},\ and\ \citenamefont {Boots}}]{doi:10.1086/653002}%
  \BibitemOpen
  \bibfield  {author} {\bibinfo {author} {\bibfnamefont {A.}~\bibnamefont {Best}}, \bibinfo {author} {\bibfnamefont {A.}~\bibnamefont {White}}, \bibinfo {author} {\bibfnamefont {E.}~\bibnamefont {Kisdi}}, \bibinfo {author} {\bibfnamefont {J.}~\bibnamefont {Antonovics}}, \bibinfo {author} {\bibfnamefont {M.}~\bibnamefont {Brockhurst}},\ and\ \bibinfo {author} {\bibfnamefont {M.}~\bibnamefont {Boots}},\ }\bibfield  {title} {\bibinfo {title} {The evolution of host‐parasite range.},\ }\href {https://doi.org/10.1086/653002} {\bibfield  {journal} {\bibinfo  {journal} {The American Naturalist}\ }\textbf {\bibinfo {volume} {176}},\ \bibinfo {pages} {63} (\bibinfo {year} {2010})},\ \bibinfo {note} {pMID: 20465424}\BibitemShut {NoStop}%
\bibitem [{\citenamefont {Hund}\ \emph {et~al.}(2022)\citenamefont {Hund}, \citenamefont {Fuess}, \citenamefont {Kenney}, \citenamefont {Maciejewski}, \citenamefont {Marini}, \citenamefont {Shim},\ and\ \citenamefont {Bolnick}}]{10.1002/evl3.274}%
  \BibitemOpen
  \bibfield  {author} {\bibinfo {author} {\bibfnamefont {A.~K.}\ \bibnamefont {Hund}}, \bibinfo {author} {\bibfnamefont {L.~E.}\ \bibnamefont {Fuess}}, \bibinfo {author} {\bibfnamefont {M.~L.}\ \bibnamefont {Kenney}}, \bibinfo {author} {\bibfnamefont {M.~F.}\ \bibnamefont {Maciejewski}}, \bibinfo {author} {\bibfnamefont {J.~M.}\ \bibnamefont {Marini}}, \bibinfo {author} {\bibfnamefont {K.~C.}\ \bibnamefont {Shim}},\ and\ \bibinfo {author} {\bibfnamefont {D.~I.}\ \bibnamefont {Bolnick}},\ }\bibfield  {title} {\bibinfo {title} {Population-level variation in parasite resistance due to differences in immune initiation and rate of response},\ }\href {https://doi.org/10.1002/evl3.274} {\bibfield  {journal} {\bibinfo  {journal} {Evolution Letters}\ }\textbf {\bibinfo {volume} {6}},\ \bibinfo {pages} {162} (\bibinfo {year} {2022})}\BibitemShut {NoStop}%
\bibitem [{\citenamefont {White}\ and\ \citenamefont {Perkins}(2012)}]{https://doi.org/10.1111/1365-2435.12012}%
  \BibitemOpen
  \bibfield  {author} {\bibinfo {author} {\bibfnamefont {T.~A.}\ \bibnamefont {White}}\ and\ \bibinfo {author} {\bibfnamefont {S.~E.}\ \bibnamefont {Perkins}},\ }\bibfield  {title} {\bibinfo {title} {The ecoimmunology of invasive species},\ }\href {https://doi.org/https://doi.org/10.1111/1365-2435.12012} {\bibfield  {journal} {\bibinfo  {journal} {Functional Ecology}\ }\textbf {\bibinfo {volume} {26}},\ \bibinfo {pages} {1313} (\bibinfo {year} {2012})}\BibitemShut {NoStop}%
\bibitem [{\citenamefont {Brown}\ \emph {et~al.}(2025)\citenamefont {Brown}, \citenamefont {Shine},\ and\ \citenamefont {Rollins}}]{Brown2025-ue}%
  \BibitemOpen
  \bibfield  {author} {\bibinfo {author} {\bibfnamefont {G.~P.}\ \bibnamefont {Brown}}, \bibinfo {author} {\bibfnamefont {R.}~\bibnamefont {Shine}},\ and\ \bibinfo {author} {\bibfnamefont {L.~A.}\ \bibnamefont {Rollins}},\ }\bibfield  {title} {\bibinfo {title} {Does a biological invasion modify host immune responses to parasite infection?},\ }\href@noop {} {\bibfield  {journal} {\bibinfo  {journal} {R Soc Open Sci}\ }\textbf {\bibinfo {volume} {12}},\ \bibinfo {pages} {240669} (\bibinfo {year} {2025})}\BibitemShut {NoStop}%
\bibitem [{\citenamefont {Jacquemyn}\ \emph {et~al.}(2003)\citenamefont {Jacquemyn}, \citenamefont {Butaye},\ and\ \citenamefont {Hermy}}]{https://doi.org/10.1111/j.0906-7590.2003.03620.x}%
  \BibitemOpen
  \bibfield  {author} {\bibinfo {author} {\bibfnamefont {H.}~\bibnamefont {Jacquemyn}}, \bibinfo {author} {\bibfnamefont {J.}~\bibnamefont {Butaye}},\ and\ \bibinfo {author} {\bibfnamefont {M.}~\bibnamefont {Hermy}},\ }\bibfield  {title} {\bibinfo {title} {Influence of environmental and spatial variables on regional distribution of forest plant species in a fragmented and changing landscape},\ }\href {https://doi.org/https://doi.org/10.1111/j.0906-7590.2003.03620.x} {\bibfield  {journal} {\bibinfo  {journal} {Ecography}\ }\textbf {\bibinfo {volume} {26}},\ \bibinfo {pages} {768} (\bibinfo {year} {2003})}\BibitemShut {NoStop}%
\bibitem [{\citenamefont {Marro}\ and\ \citenamefont {Dickman}(1999)}]{MarroDickman1999}%
  \BibitemOpen
  \bibfield  {author} {\bibinfo {author} {\bibfnamefont {J.}~\bibnamefont {Marro}}\ and\ \bibinfo {author} {\bibfnamefont {R.}~\bibnamefont {Dickman}},\ }\bibfield  {title} {\bibinfo {title} {Nonequilibrium phase transitions in lattice models}\ }(\bibinfo {year} {1999})\BibitemShut {NoStop}%
\bibitem [{\citenamefont {Hinrichsen}(2000)}]{Hinrichsen2000}%
  \BibitemOpen
  \bibfield  {author} {\bibinfo {author} {\bibfnamefont {H.}~\bibnamefont {Hinrichsen}},\ }\bibfield  {title} {\bibinfo {title} {Non-equilibrium critical phenomena and phase transitions into absorbing states},\ }\href {https://doi.org/10.1080/00018730050198152} {\bibfield  {journal} {\bibinfo  {journal} {Advances in Physics}\ }\textbf {\bibinfo {volume} {49}},\ \bibinfo {pages} {815} (\bibinfo {year} {2000})}\BibitemShut {NoStop}%
\bibitem [{\citenamefont {Henkel}\ \emph {et~al.}(2008)\citenamefont {Henkel}, \citenamefont {Hinrichsen},\ and\ \citenamefont {L{\"u}beck}}]{henkel2008non}%
  \BibitemOpen
  \bibfield  {author} {\bibinfo {author} {\bibfnamefont {M.}~\bibnamefont {Henkel}}, \bibinfo {author} {\bibfnamefont {H.}~\bibnamefont {Hinrichsen}},\ and\ \bibinfo {author} {\bibfnamefont {S.}~\bibnamefont {L{\"u}beck}},\ }\href {https://books.google.hu/books?id=OKbtkq4A-1EC} {\emph {\bibinfo {title} {Non-Equilibrium Phase Transitions: Volume 1: Absorbing Phase Transitions}}},\ Theoretical and Mathematical Physics\ (\bibinfo  {publisher} {Springer Netherlands},\ \bibinfo {year} {2008})\BibitemShut {NoStop}%
\bibitem [{\citenamefont {\'Odor}(2004)}]{RevModPhys.76.663}%
  \BibitemOpen
  \bibfield  {author} {\bibinfo {author} {\bibfnamefont {G.}~\bibnamefont {\'Odor}},\ }\bibfield  {title} {\bibinfo {title} {Universality classes in nonequilibrium lattice systems},\ }\href {https://doi.org/10.1103/RevModPhys.76.663} {\bibfield  {journal} {\bibinfo  {journal} {Rev. Mod. Phys.}\ }\textbf {\bibinfo {volume} {76}},\ \bibinfo {pages} {663} (\bibinfo {year} {2004})}\BibitemShut {NoStop}%
\bibitem [{\citenamefont {Janssen}(1981)}]{Janssen1981}%
  \BibitemOpen
  \bibfield  {author} {\bibinfo {author} {\bibfnamefont {H.~K.}\ \bibnamefont {Janssen}},\ }\bibfield  {title} {\bibinfo {title} {On the nonequilibrium phase transition in reaction-diffusion systems with an absorbing stationary state},\ }\href {https://doi.org/10.1007/BF01319549} {\bibfield  {journal} {\bibinfo  {journal} {Zeitschrift für Physik B Condensed Matter}\ }\textbf {\bibinfo {volume} {42}},\ \bibinfo {pages} {151} (\bibinfo {year} {1981})}\BibitemShut {NoStop}%
\bibitem [{\citenamefont {Grassberger}(1982)}]{Grassberger1982}%
  \BibitemOpen
  \bibfield  {author} {\bibinfo {author} {\bibfnamefont {P.}~\bibnamefont {Grassberger}},\ }\bibfield  {title} {\bibinfo {title} {On phase transitions in {S}chlögl's second model},\ }\href {https://doi.org/10.1007/BF01313803} {\bibfield  {journal} {\bibinfo  {journal} {Zeitschrift für Physik B Condensed Matter}\ }\textbf {\bibinfo {volume} {47}},\ \bibinfo {pages} {365} (\bibinfo {year} {1982})}\BibitemShut {NoStop}%
\bibitem [{\citenamefont {Ziff}\ \emph {et~al.}(1986)\citenamefont {Ziff}, \citenamefont {Gulari},\ and\ \citenamefont {Barshad}}]{ZiffGulariBarshad1986}%
  \BibitemOpen
  \bibfield  {author} {\bibinfo {author} {\bibfnamefont {R.~M.}\ \bibnamefont {Ziff}}, \bibinfo {author} {\bibfnamefont {E.}~\bibnamefont {Gulari}},\ and\ \bibinfo {author} {\bibfnamefont {Y.}~\bibnamefont {Barshad}},\ }\bibfield  {title} {\bibinfo {title} {Kinetic phase transitions in an irreversible surface-reaction model},\ }\href {https://doi.org/10.1103/PhysRevLett.56.2553} {\bibfield  {journal} {\bibinfo  {journal} {Phys. Rev. Lett.}\ }\textbf {\bibinfo {volume} {56}},\ \bibinfo {pages} {2553} (\bibinfo {year} {1986})}\BibitemShut {NoStop}%
\bibitem [{\citenamefont {Grassberger}(1995)}]{Grassberger1995}%
  \BibitemOpen
  \bibfield  {author} {\bibinfo {author} {\bibfnamefont {P.}~\bibnamefont {Grassberger}},\ }\bibfield  {title} {\bibinfo {title} {Are damage spreading transitions generically in the universality class of directed percolation?},\ }\href {https://doi.org/10.1007/BF02179381} {\bibfield  {journal} {\bibinfo  {journal} {Journal of Statistical Physics}\ }\textbf {\bibinfo {volume} {79}},\ \bibinfo {pages} {13} (\bibinfo {year} {1995})}\BibitemShut {NoStop}%
\bibitem [{\citenamefont {Tang}\ and\ \citenamefont {Leschhorn}(1992)}]{TangLeschorn1992}%
  \BibitemOpen
  \bibfield  {author} {\bibinfo {author} {\bibfnamefont {L.-H.}\ \bibnamefont {Tang}}\ and\ \bibinfo {author} {\bibfnamefont {H.}~\bibnamefont {Leschhorn}},\ }\bibfield  {title} {\bibinfo {title} {Pinning by directed percolation},\ }\href {https://doi.org/10.1103/PhysRevA.45.R8309} {\bibfield  {journal} {\bibinfo  {journal} {Phys. Rev. A}\ }\textbf {\bibinfo {volume} {45}},\ \bibinfo {pages} {R8309} (\bibinfo {year} {1992})}\BibitemShut {NoStop}%
\bibitem [{\citenamefont {Sch\"utz}(2000)}]{schutz}%
  \BibitemOpen
  \bibfield  {author} {\bibinfo {author} {\bibfnamefont {G.~M.}\ \bibnamefont {Sch\"utz}},\ }\bibfield  {title} {\bibinfo {title} {Exactly solvable models for many-body systems far from equilibrium},\ }in\ \href@noop {} {\emph {\bibinfo {booktitle} {Phase Transitions and Critical Phenomena}}},\ Vol.~\bibinfo {volume} {19},\ \bibinfo {editor} {edited by\ \bibinfo {editor} {\bibfnamefont {C.}~\bibnamefont {Domb}}\ and\ \bibinfo {editor} {\bibfnamefont {J.~L.}\ \bibnamefont {Lebowitz}}}\ (\bibinfo  {publisher} {Academic, London},\ \bibinfo {year} {2000})\BibitemShut {NoStop}%
\bibitem [{\citenamefont {J.~Hooyberghs}(2001)}]{vanderzande}%
  \BibitemOpen
  \bibfield  {author} {\bibinfo {author} {\bibfnamefont {C.~V.}\ \bibnamefont {J.~Hooyberghs}},\ }\bibfield  {title} {\bibinfo {title} {One-dimensional contact process: duality and renormalization},\ }\href@noop {} {\bibfield  {journal} {\bibinfo  {journal} {Phys. Rev. E}\ }\textbf {\bibinfo {volume} {63}},\ \bibinfo {pages} {041109} (\bibinfo {year} {2001})}\BibitemShut {NoStop}%
\bibitem [{\citenamefont {Dickman}(1999)}]{Dickman1999}%
  \BibitemOpen
  \bibfield  {author} {\bibinfo {author} {\bibfnamefont {R.}~\bibnamefont {Dickman}},\ }\bibfield  {title} {\bibinfo {title} {Reweighting in nonequilibrium simulations},\ }\href {https://doi.org/10.1103/PhysRevE.60.R2441} {\bibfield  {journal} {\bibinfo  {journal} {Phys. Rev. E}\ }\textbf {\bibinfo {volume} {60}},\ \bibinfo {pages} {R2441} (\bibinfo {year} {1999})}\BibitemShut {NoStop}%
\bibitem [{\citenamefont {Miller}\ and\ \citenamefont {MacKay}(1994)}]{Miller}%
  \BibitemOpen
  \bibfield  {author} {\bibinfo {author} {\bibfnamefont {K.~D.}\ \bibnamefont {Miller}}\ and\ \bibinfo {author} {\bibfnamefont {D.~J.~C.}\ \bibnamefont {MacKay}},\ }\bibfield  {title} {\bibinfo {title} {The role of constraints in {H}ebbian learning},\ }\href {https://doi.org/10.1162/neco.1994.6.1.100} {\bibfield  {journal} {\bibinfo  {journal} {Neural Computation}\ }\textbf {\bibinfo {volume} {6}},\ \bibinfo {pages} {100} (\bibinfo {year} {1994})}\BibitemShut {NoStop}%
\bibitem [{\citenamefont {Gross}\ \emph {et~al.}(2006)\citenamefont {Gross}, \citenamefont {D'Lima},\ and\ \citenamefont {Blasius}}]{PhysRevLett.96.208701}%
  \BibitemOpen
  \bibfield  {author} {\bibinfo {author} {\bibfnamefont {T.}~\bibnamefont {Gross}}, \bibinfo {author} {\bibfnamefont {C.~J.~D.}\ \bibnamefont {D'Lima}},\ and\ \bibinfo {author} {\bibfnamefont {B.}~\bibnamefont {Blasius}},\ }\bibfield  {title} {\bibinfo {title} {Epidemic dynamics on an adaptive network},\ }\href {https://doi.org/10.1103/PhysRevLett.96.208701} {\bibfield  {journal} {\bibinfo  {journal} {Phys. Rev. Lett.}\ }\textbf {\bibinfo {volume} {96}},\ \bibinfo {pages} {208701} (\bibinfo {year} {2006})}\BibitemShut {NoStop}%
\bibitem [{\citenamefont {Gross}\ and\ \citenamefont {Sayama}(2009)}]{adaptive}%
  \BibitemOpen
  \bibfield  {author} {\bibinfo {author} {\bibfnamefont {T.}~\bibnamefont {Gross}}\ and\ \bibinfo {author} {\bibfnamefont {H.}~\bibnamefont {Sayama}},\ }\href@noop {} {\emph {\bibinfo {title} {Adaptive Networks. Theory, Models and Applications}}}\ (\bibinfo  {publisher} {Springer Verlag},\ \bibinfo {address} {Germany},\ \bibinfo {year} {2009})\BibitemShut {NoStop}%
\bibitem [{\citenamefont {{\'O}dor}\ and\ \citenamefont {Karsai}(2025)}]{Odor2025}%
  \BibitemOpen
  \bibfield  {author} {\bibinfo {author} {\bibfnamefont {G.}~\bibnamefont {{\'O}dor}}\ and\ \bibinfo {author} {\bibfnamefont {M.}~\bibnamefont {Karsai}},\ }\bibfield  {title} {\bibinfo {title} {Epidemic-induced local awareness behavior inferred from surveys and genetic sequence data},\ }\href {https://doi.org/10.1038/s41467-025-59508-5} {\bibfield  {journal} {\bibinfo  {journal} {Nature Communications}\ }\textbf {\bibinfo {volume} {16}},\ \bibinfo {pages} {4758} (\bibinfo {year} {2025})}\BibitemShut {NoStop}%
\bibitem [{\citenamefont {Kolok}\ \emph {et~al.}(2025)\citenamefont {Kolok}, \citenamefont {\'Odor}, \citenamefont {Keliger},\ and\ \citenamefont {Karsai}}]{PhysRevResearch.7.L012061}%
  \BibitemOpen
  \bibfield  {author} {\bibinfo {author} {\bibfnamefont {C.~B.}\ \bibnamefont {Kolok}}, \bibinfo {author} {\bibfnamefont {G.}~\bibnamefont {\'Odor}}, \bibinfo {author} {\bibfnamefont {D.}~\bibnamefont {Keliger}},\ and\ \bibinfo {author} {\bibfnamefont {M.}~\bibnamefont {Karsai}},\ }\bibfield  {title} {\bibinfo {title} {Epidemic paradox induced by awareness driven network dynamics},\ }\href {https://doi.org/10.1103/PhysRevResearch.7.L012061} {\bibfield  {journal} {\bibinfo  {journal} {Phys. Rev. Res.}\ }\textbf {\bibinfo {volume} {7}},\ \bibinfo {pages} {L012061} (\bibinfo {year} {2025})}\BibitemShut {NoStop}%
\bibitem [{\citenamefont {Jensen}\ and\ \citenamefont {Dickman}(1993)}]{JensenDickman1993}%
  \BibitemOpen
  \bibfield  {author} {\bibinfo {author} {\bibfnamefont {I.}~\bibnamefont {Jensen}}\ and\ \bibinfo {author} {\bibfnamefont {R.}~\bibnamefont {Dickman}},\ }\bibfield  {title} {\bibinfo {title} {Time-dependent perturbation theory for nonequilibrium lattice models},\ }\href {https://doi.org/10.1007/bf01048090} {\bibfield  {journal} {\bibinfo  {journal} {Journal of Statistical Physics}\ }\textbf {\bibinfo {volume} {71}},\ \bibinfo {pages} {89–127} (\bibinfo {year} {1993})}\BibitemShut {NoStop}%
\bibitem [{\citenamefont {Jensen}(1999)}]{Jensen1999}%
  \BibitemOpen
  \bibfield  {author} {\bibinfo {author} {\bibfnamefont {I.}~\bibnamefont {Jensen}},\ }\bibfield  {title} {\bibinfo {title} {Low-density series expansions for directed percolation: I. a new efficient algorithm with applications to the square lattice},\ }\href {https://doi.org/10.1088/0305-4470/32/28/304} {\bibfield  {journal} {\bibinfo  {journal} {Journal of Physics A: Mathematical and General}\ }\textbf {\bibinfo {volume} {32}},\ \bibinfo {pages} {5233–5249} (\bibinfo {year} {1999})}\BibitemShut {NoStop}%
\bibitem [{\citenamefont {Vojta}\ \emph {et~al.}(2009)\citenamefont {Vojta}, \citenamefont {Farquhar},\ and\ \citenamefont {Mast}}]{VojtaFarquharMast2009}%
  \BibitemOpen
  \bibfield  {author} {\bibinfo {author} {\bibfnamefont {T.}~\bibnamefont {Vojta}}, \bibinfo {author} {\bibfnamefont {A.}~\bibnamefont {Farquhar}},\ and\ \bibinfo {author} {\bibfnamefont {J.}~\bibnamefont {Mast}},\ }\bibfield  {title} {\bibinfo {title} {Infinite-randomness critical point in the two-dimensional disordered contact process},\ }\href {https://doi.org/10.1103/PhysRevE.79.011111} {\bibfield  {journal} {\bibinfo  {journal} {Phys. Rev. E}\ }\textbf {\bibinfo {volume} {79}},\ \bibinfo {pages} {011111} (\bibinfo {year} {2009})}\BibitemShut {NoStop}%
\bibitem [{\citenamefont {Tom\'e}\ and\ \citenamefont {Ziff}(2010)}]{Tome}%
  \BibitemOpen
  \bibfield  {author} {\bibinfo {author} {\bibfnamefont {T.}~\bibnamefont {Tom\'e}}\ and\ \bibinfo {author} {\bibfnamefont {R.~M.}\ \bibnamefont {Ziff}},\ }\bibfield  {title} {\bibinfo {title} {Critical behavior of the susceptible-infected-recovered model on a square lattice},\ }\href {https://doi.org/10.1103/PhysRevE.82.051921} {\bibfield  {journal} {\bibinfo  {journal} {Phys. Rev. E}\ }\textbf {\bibinfo {volume} {82}},\ \bibinfo {pages} {051921} (\bibinfo {year} {2010})}\BibitemShut {NoStop}%
\bibitem [{\citenamefont {Mu\~noz}\ \emph {et~al.}(1999)\citenamefont {Mu\~noz}, \citenamefont {Dickman}, \citenamefont {Vespignani},\ and\ \citenamefont {Zapperi}}]{PhysRevE.59.6175}%
  \BibitemOpen
  \bibfield  {author} {\bibinfo {author} {\bibfnamefont {M.~A.}\ \bibnamefont {Mu\~noz}}, \bibinfo {author} {\bibfnamefont {R.}~\bibnamefont {Dickman}}, \bibinfo {author} {\bibfnamefont {A.}~\bibnamefont {Vespignani}},\ and\ \bibinfo {author} {\bibfnamefont {S.}~\bibnamefont {Zapperi}},\ }\bibfield  {title} {\bibinfo {title} {Avalanche and spreading exponents in systems with absorbing states},\ }\href {https://doi.org/10.1103/PhysRevE.59.6175} {\bibfield  {journal} {\bibinfo  {journal} {Phys. Rev. E}\ }\textbf {\bibinfo {volume} {59}},\ \bibinfo {pages} {6175} (\bibinfo {year} {1999})}\BibitemShut {NoStop}%
\bibitem [{\citenamefont {Tretyakov A~Yu}(1997)}]{tretyakov}%
  \BibitemOpen
  \bibfield  {author} {\bibinfo {author} {\bibfnamefont {K.~N.}\ \bibnamefont {Tretyakov A~Yu}, \bibfnamefont {Inui~N}},\ }\bibfield  {title} {\bibinfo {title} {Phase transition for the one-sided contact process},\ }\href@noop {} {\bibfield  {journal} {\bibinfo  {journal} {J. Phys. Soc. Jap.}\ }\textbf {\bibinfo {volume} {66}},\ \bibinfo {pages} {3764} (\bibinfo {year} {1997})}\BibitemShut {NoStop}%
\bibitem [{\citenamefont {Delsuc}(2003)}]{Delsuc}%
  \BibitemOpen
  \bibfield  {author} {\bibinfo {author} {\bibfnamefont {F.}~\bibnamefont {Delsuc}},\ }\bibfield  {title} {\bibinfo {title} {Army ants trapped by their evolutionary history},\ }\href {https://doi.org/10.1371/journal.pbio.0000037} {\bibfield  {journal} {\bibinfo  {journal} {PLOS Biology}\ }\textbf {\bibinfo {volume} {1}},\ \bibinfo {pages} {null} (\bibinfo {year} {2003})}\BibitemShut {NoStop}%
\bibitem [{\citenamefont {Erhard}\ \emph {et~al.}(2022)\citenamefont {Erhard}, \citenamefont {Franco},\ and\ \citenamefont {Reis}}]{Erhard2022}%
  \BibitemOpen
  \bibfield  {author} {\bibinfo {author} {\bibfnamefont {D.}~\bibnamefont {Erhard}}, \bibinfo {author} {\bibfnamefont {T.}~\bibnamefont {Franco}},\ and\ \bibinfo {author} {\bibfnamefont {G.}~\bibnamefont {Reis}},\ }\bibfield  {title} {\bibinfo {title} {The directed edge reinforced random walk: The ant mill phenomenon},\ }\href {https://doi.org/10.1007/s10955-022-03031-0} {\bibfield  {journal} {\bibinfo  {journal} {Journal of Statistical Physics}\ }\textbf {\bibinfo {volume} {190}},\ \bibinfo {pages} {18} (\bibinfo {year} {2022})}\BibitemShut {NoStop}%
\bibitem [{\citenamefont {Iglói}\ and\ \citenamefont {Monthus}(2005)}]{IgloiMonthus2005}%
  \BibitemOpen
  \bibfield  {author} {\bibinfo {author} {\bibfnamefont {F.}~\bibnamefont {Iglói}}\ and\ \bibinfo {author} {\bibfnamefont {C.}~\bibnamefont {Monthus}},\ }\bibfield  {title} {\bibinfo {title} {Strong disorder {RG} approach of random systems},\ }\href {https://doi.org/https://doi.org/10.1016/j.physrep.2005.02.006} {\bibfield  {journal} {\bibinfo  {journal} {Physics Reports}\ }\textbf {\bibinfo {volume} {412}},\ \bibinfo {pages} {277} (\bibinfo {year} {2005})}\BibitemShut {NoStop}%
\bibitem [{\citenamefont {Igl{\'o}i}\ and\ \citenamefont {Monthus}(2018)}]{Igloi2018}%
  \BibitemOpen
  \bibfield  {author} {\bibinfo {author} {\bibfnamefont {F.}~\bibnamefont {Igl{\'o}i}}\ and\ \bibinfo {author} {\bibfnamefont {C.}~\bibnamefont {Monthus}},\ }\bibfield  {title} {\bibinfo {title} {Strong disorder {RG} approach -- a short review of recent developments},\ }\href {https://doi.org/10.1140/epjb/e2018-90434-8} {\bibfield  {journal} {\bibinfo  {journal} {The European Physical Journal B}\ }\textbf {\bibinfo {volume} {91}},\ \bibinfo {pages} {290} (\bibinfo {year} {2018})}\BibitemShut {NoStop}%
\bibitem [{\citenamefont {Buend{\'\i}a}\ \emph {et~al.}(2022)\citenamefont {Buend{\'\i}a}, \citenamefont {Villegas}, \citenamefont {Burioni},\ and\ \citenamefont {Mu{\~n}oz}}]{buendia2022broad}%
  \BibitemOpen
  \bibfield  {author} {\bibinfo {author} {\bibfnamefont {V.}~\bibnamefont {Buend{\'\i}a}}, \bibinfo {author} {\bibfnamefont {P.}~\bibnamefont {Villegas}}, \bibinfo {author} {\bibfnamefont {R.}~\bibnamefont {Burioni}},\ and\ \bibinfo {author} {\bibfnamefont {M.~A.}\ \bibnamefont {Mu{\~n}oz}},\ }\bibfield  {title} {\bibinfo {title} {The broad edge of synchronization: Griffiths effects and collective phenomena in brain networks},\ }\href@noop {} {\bibfield  {journal} {\bibinfo  {journal} {Philosophical Transactions of the Royal Society A}\ }\textbf {\bibinfo {volume} {380}},\ \bibinfo {pages} {20200424} (\bibinfo {year} {2022})}\BibitemShut {NoStop}%
\bibitem [{\citenamefont {Girardi-Schappo}\ \emph {et~al.}(2016)\citenamefont {Girardi-Schappo}, \citenamefont {Bortolotto}, \citenamefont {Gonsalves}, \citenamefont {Pinto},\ and\ \citenamefont {Tragtenberg}}]{Girardi-Schappo2016}%
  \BibitemOpen
  \bibfield  {author} {\bibinfo {author} {\bibfnamefont {M.}~\bibnamefont {Girardi-Schappo}}, \bibinfo {author} {\bibfnamefont {G.~S.}\ \bibnamefont {Bortolotto}}, \bibinfo {author} {\bibfnamefont {J.~J.}\ \bibnamefont {Gonsalves}}, \bibinfo {author} {\bibfnamefont {L.~T.}\ \bibnamefont {Pinto}},\ and\ \bibinfo {author} {\bibfnamefont {M.~H.~R.}\ \bibnamefont {Tragtenberg}},\ }\bibfield  {title} {\bibinfo {title} {Griffiths phase and long-range correlations in a biologically motivated visual cortex model},\ }\href {https://doi.org/10.1038/srep29561} {\bibfield  {journal} {\bibinfo  {journal} {Scientific Reports}\ }\textbf {\bibinfo {volume} {6}},\ \bibinfo {pages} {29561} (\bibinfo {year} {2016})}\BibitemShut {NoStop}%
\bibitem [{\citenamefont {Pretel}\ \emph {et~al.}(2024)\citenamefont {Pretel}, \citenamefont {Buend\'{\i}a}, \citenamefont {Torres},\ and\ \citenamefont {Mu\~noz}}]{PhysRevResearch.6.023018}%
  \BibitemOpen
  \bibfield  {author} {\bibinfo {author} {\bibfnamefont {J.}~\bibnamefont {Pretel}}, \bibinfo {author} {\bibfnamefont {V.}~\bibnamefont {Buend\'{\i}a}}, \bibinfo {author} {\bibfnamefont {J.~J.}\ \bibnamefont {Torres}},\ and\ \bibinfo {author} {\bibfnamefont {M.~A.}\ \bibnamefont {Mu\~noz}},\ }\bibfield  {title} {\bibinfo {title} {From asynchronous states to griffiths phases and back: Structural heterogeneity and homeostasis in excitatory-inhibitory networks},\ }\href {https://doi.org/10.1103/PhysRevResearch.6.023018} {\bibfield  {journal} {\bibinfo  {journal} {Phys. Rev. Res.}\ }\textbf {\bibinfo {volume} {6}},\ \bibinfo {pages} {023018} (\bibinfo {year} {2024})}\BibitemShut {NoStop}%
\bibitem [{\citenamefont {Fusc{\`a}}\ \emph {et~al.}(2023)\citenamefont {Fusc{\`a}}, \citenamefont {Siebenh{\"u}hner}, \citenamefont {Wang}, \citenamefont {Myrov}, \citenamefont {Arnulfo}, \citenamefont {Nobili}, \citenamefont {Palva},\ and\ \citenamefont {Palva}}]{Fusca2023}%
  \BibitemOpen
  \bibfield  {author} {\bibinfo {author} {\bibfnamefont {M.}~\bibnamefont {Fusc{\`a}}}, \bibinfo {author} {\bibfnamefont {F.}~\bibnamefont {Siebenh{\"u}hner}}, \bibinfo {author} {\bibfnamefont {S.~H.}\ \bibnamefont {Wang}}, \bibinfo {author} {\bibfnamefont {V.}~\bibnamefont {Myrov}}, \bibinfo {author} {\bibfnamefont {G.}~\bibnamefont {Arnulfo}}, \bibinfo {author} {\bibfnamefont {L.}~\bibnamefont {Nobili}}, \bibinfo {author} {\bibfnamefont {J.~M.}\ \bibnamefont {Palva}},\ and\ \bibinfo {author} {\bibfnamefont {S.}~\bibnamefont {Palva}},\ }\bibfield  {title} {\bibinfo {title} {Brain criticality predicts individual levels of inter-areal synchronization in human electrophysiological data},\ }\href {https://doi.org/10.1038/s41467-023-40056-9} {\bibfield  {journal} {\bibinfo  {journal} {Nature Communications}\ }\textbf {\bibinfo {volume} {14}},\ \bibinfo {pages} {4736} (\bibinfo {year} {2023})}\BibitemShut {NoStop}%
\bibitem [{\citenamefont {{\'O}dor}\ \emph {et~al.}(2015)\citenamefont {{\'O}dor}, \citenamefont {Dickman},\ and\ \citenamefont {{\'O}dor}}]{Odor2015}%
  \BibitemOpen
  \bibfield  {author} {\bibinfo {author} {\bibfnamefont {G.}~\bibnamefont {{\'O}dor}}, \bibinfo {author} {\bibfnamefont {R.}~\bibnamefont {Dickman}},\ and\ \bibinfo {author} {\bibfnamefont {G.}~\bibnamefont {{\'O}dor}},\ }\bibfield  {title} {\bibinfo {title} {Griffiths phases and localization in hierarchical modular networks},\ }\href {https://doi.org/10.1038/srep14451} {\bibfield  {journal} {\bibinfo  {journal} {Scientific Reports}\ }\textbf {\bibinfo {volume} {5}},\ \bibinfo {pages} {14451} (\bibinfo {year} {2015})}\BibitemShut {NoStop}%
\bibitem [{\citenamefont {{Wright}}\ \emph {et~al.}(2022)\citenamefont {{Wright}}, \citenamefont {{Anderson}},\ and\ \citenamefont {{McMahon}}}]{Logan}%
  \BibitemOpen
  \bibfield  {author} {\bibinfo {author} {\bibfnamefont {L.}~\bibnamefont {{Wright}}}, \bibinfo {author} {\bibfnamefont {M.}~\bibnamefont {{Anderson}}},\ and\ \bibinfo {author} {\bibfnamefont {P.}~\bibnamefont {{McMahon}}},\ }\bibfield  {title} {\bibinfo {title} {Evidence for {G}riffiths phase criticality in residual neural networks},\ }in\ \href@noop {} {\emph {\bibinfo {booktitle} {APS March Meeting Abstracts}}},\ \bibinfo {series} {APS Meeting Abstracts}, Vol.\ \bibinfo {volume} {2022}\ (\bibinfo {year} {2022})\ p.\ \bibinfo {pages} {M09.012}\BibitemShut {NoStop}%
\bibitem [{\citenamefont {Vojta}(2012)}]{Vojta2012}%
  \BibitemOpen
  \bibfield  {author} {\bibinfo {author} {\bibfnamefont {T.}~\bibnamefont {Vojta}},\ }\bibfield  {title} {\bibinfo {title} {Monte {C}arlo simulations of the clean and disordered contact process in three dimensions},\ }\href {https://doi.org/10.1103/PhysRevE.86.051137} {\bibfield  {journal} {\bibinfo  {journal} {Phys. Rev. E}\ }\textbf {\bibinfo {volume} {86}},\ \bibinfo {pages} {051137} (\bibinfo {year} {2012})}\BibitemShut {NoStop}%
\bibitem [{\citenamefont {Juhász}\ and\ \citenamefont {Kovács}(2013)}]{Juhasz_2013}%
  \BibitemOpen
  \bibfield  {author} {\bibinfo {author} {\bibfnamefont {R.}~\bibnamefont {Juhász}}\ and\ \bibinfo {author} {\bibfnamefont {I.~A.}\ \bibnamefont {Kovács}},\ }\bibfield  {title} {\bibinfo {title} {Infinite randomness critical behavior of the contact process on networks with long-range connections},\ }\href {https://doi.org/10.1088/1742-5468/2013/06/P06003} {\bibfield  {journal} {\bibinfo  {journal} {Journal of Statistical Mechanics: Theory and Experiment}\ }\textbf {\bibinfo {volume} {2013}},\ \bibinfo {pages} {P06003} (\bibinfo {year} {2013})}\BibitemShut {NoStop}%
\bibitem [{\citenamefont {Harris}(1974{\natexlab{b}})}]{Harris1974criterion}%
  \BibitemOpen
  \bibfield  {author} {\bibinfo {author} {\bibfnamefont {A.~B.}\ \bibnamefont {Harris}},\ }\bibfield  {title} {\bibinfo {title} {Effect of random defects on the critical behaviour of {I}sing models},\ }\href {https://doi.org/10.1088/0022-3719/7/9/009} {\bibfield  {journal} {\bibinfo  {journal} {Journal of Physics C: Solid State Physics}\ }\textbf {\bibinfo {volume} {7}},\ \bibinfo {pages} {1671} (\bibinfo {year} {1974}{\natexlab{b}})}\BibitemShut {NoStop}%
\bibitem [{\citenamefont {Castellano}\ and\ \citenamefont {Pastor-Satorras}(2006)}]{PhysRevLett.96.038701}%
  \BibitemOpen
  \bibfield  {author} {\bibinfo {author} {\bibfnamefont {C.}~\bibnamefont {Castellano}}\ and\ \bibinfo {author} {\bibfnamefont {R.}~\bibnamefont {Pastor-Satorras}},\ }\bibfield  {title} {\bibinfo {title} {Non-mean-field behavior of the contact process on scale-free networks},\ }\href {https://doi.org/10.1103/PhysRevLett.96.038701} {\bibfield  {journal} {\bibinfo  {journal} {Phys. Rev. Lett.}\ }\textbf {\bibinfo {volume} {96}},\ \bibinfo {pages} {038701} (\bibinfo {year} {2006})}\BibitemShut {NoStop}%
\bibitem [{\citenamefont {Pastor-Satorras}\ and\ \citenamefont {Vespignani}(2001)}]{PhysRevE.63.066117}%
  \BibitemOpen
  \bibfield  {author} {\bibinfo {author} {\bibfnamefont {R.}~\bibnamefont {Pastor-Satorras}}\ and\ \bibinfo {author} {\bibfnamefont {A.}~\bibnamefont {Vespignani}},\ }\bibfield  {title} {\bibinfo {title} {Epidemic dynamics and endemic states in complex networks},\ }\href {https://doi.org/10.1103/PhysRevE.63.066117} {\bibfield  {journal} {\bibinfo  {journal} {Phys. Rev. E}\ }\textbf {\bibinfo {volume} {63}},\ \bibinfo {pages} {066117} (\bibinfo {year} {2001})}\BibitemShut {NoStop}%
\bibitem [{\citenamefont {Castellano}\ and\ \citenamefont {Pastor-Satorras}(2010)}]{PhysRevLett.105.218701}%
  \BibitemOpen
  \bibfield  {author} {\bibinfo {author} {\bibfnamefont {C.}~\bibnamefont {Castellano}}\ and\ \bibinfo {author} {\bibfnamefont {R.}~\bibnamefont {Pastor-Satorras}},\ }\bibfield  {title} {\bibinfo {title} {Thresholds for epidemic spreading in networks},\ }\href {https://doi.org/10.1103/PhysRevLett.105.218701} {\bibfield  {journal} {\bibinfo  {journal} {Phys. Rev. Lett.}\ }\textbf {\bibinfo {volume} {105}},\ \bibinfo {pages} {218701} (\bibinfo {year} {2010})}\BibitemShut {NoStop}%
\bibitem [{\citenamefont {Chatterjee}\ and\ \citenamefont {Durrett}(2009)}]{Durrett}%
  \BibitemOpen
  \bibfield  {author} {\bibinfo {author} {\bibfnamefont {S.}~\bibnamefont {Chatterjee}}\ and\ \bibinfo {author} {\bibfnamefont {R.}~\bibnamefont {Durrett}},\ }\bibfield  {title} {\bibinfo {title} {Contact processes on random graphs with power law degree distributions have critical value 0},\ }\href {http://www.jstor.org/stable/27795079} {\bibfield  {journal} {\bibinfo  {journal} {The Annals of Probability}\ }\textbf {\bibinfo {volume} {37}},\ \bibinfo {pages} {2332} (\bibinfo {year} {2009})}\BibitemShut {NoStop}%
\end{thebibliography}%


\end{document}